\documentclass[authoryear]{elshcj}
\usepackage{amssymb,amscd,amsmath}
\usepackage{graphicx}
\usepackage{color}
\def\hsp{\hspace}

\def\tm{truemm}
\def\hsa{\mbox{} \hsp{1mm} }
\def\hsb{\mbox{} \hsp{5mm} }
\def\hsc{\mbox{} \hsp{10mm} } 
\def\hsd{\mbox{} \hsp{15mm} }

\def\hsg{\mbox{} \hsp{-2mm} } 
\def\hsbb{\hsp{-6mm}}
\def\half{\frac{1}{2}} 
\def\beqa{\begin{eqnarray}}
\def\eeqa{\end{eqnarray}}
\def\beqan{\begin{eqnarray*}}
\def\eeqan{\end{eqnarray*}}
\def\a={&=&}
\def\aa={&\approx&}
\def\ap={&\propto&}
\def\b={\ = \ }
\def\sa={\hsg &=& \hsg}
\def\sap={\hsg &\propto&\hsg}
\def\saa={\hsg &\approx&\hsg}
\def\nnn{\nonumber \\}
\def\bit{\begin{itemize}}
\def\eit{\end{itemize}}
\def\mf{\hsp{5mm} \mbox{for} \ \ }
\def\aii{a_{ii}}
\def\aij{a_{ij}}
\def\aik{a_{ik}}
\def\aji{a_{ji}}
\def\ajj{a_{jj}}
\def\ajk{a_{jk}}
\def\aki{a_{ki}}
\def\akj{a_{kj}}
\def\akk{a_{kk}}
\def\alii{\alpha_{ii}}
\def\alij{\alpha_{ij}}
\def\alji{\alpha_{ji}}
\def\aljj{\alpha_{jj}}
\def\dcd{d_{CD}}
\def\dcl{d_{CL}}
\def\ddc{d_{DC}}
\def\ddl{d_{DL}}
\def\dlc{d_{LC}}
\def\dld{d_{LD}}
\def\dij{d_{ij}}
\def\dik{d_{ik}}

\def\dijn{d_{12}}
\def\dikn{d_{13}}
\def\djin{d_{21}}
\def\djkn{d_{23}}
\def\dkin{d_{31}}
\def\dkjn{d_{32}}
\def\fij{f_{ij}}
\def\fji{f_{ji}}
\def\fki{f_{ki}}
\def\fkj{f_{kj}}
\def\gij{g_{ij}}
\def\gji{g_{ji}}
\def\gki{g_{ki}}
\def\gkj{g_{kj}}
\def\lcd{l_{CD}}
\def\lcl{l_{CL}}

\def\ldl{l_{DL}}
\def\llc{l_{LC}}
\def\lld{l_{LD}}
\def\lij{l_{ij}}
\def\lik{l_{ik}}
\def\lji{l_{ji}}
\def\ljk{l_{jk}}
\def\lki{l_{ki}}
\def\lkj{l_{kj}}
\def\na{{N_{\!A}}}
\def\pna{\phi_{_{\!N_{\!A}}}}
\def\rab{\rho_{\!_{AB}}}
\def\rba{\rho_{\!_{BA}}}
\def\rcd{\rho_{\!_{CD}}}
\def\rcl{\rho_{\!_{CL}}}

\def\rclf{\rho_{\!_{CL_1}}}
\def\rcln{\rho_{\!_{CL_n}}}
\def\rdc{\rho_{\!_{DC}}}
\def\rdl{\rho_{\!_{DL}}}

\def\rdlf{\rho_{\!_{DL_1}}}
\def\rdln{\rho_{\!_{DL_n}}}
\def\rlc{\rho_{\!_{LC}}}
\def\rld{\rho_{\!_{LD}}}

\def\rlfc{\rho_{\!_{L_1 C}}}
\def\rlnc{\rho_{\!_{L_n C}}}

\def\rlfd{\rho_{\!_{L_1 D}}}
\def\rlnd{\rho_{\!_{L_n D}}}

\def\rlfln{\rho_{\!_{L_1 L_n}}}

\def\rlnlf{\rho_{\!_{L_n L_1}}}
\def\rxl{\rho_{\!_{iL}}}
\def\rxlj{\rho_{\!_{iL_j}}}
\def\rij{\rho_{ij}}
\def\rik{\rho_{ik}}
\def\rji{\rho_{ji}}
\def\rjk{\rho_{jk}}
\def\rki{\rho_{ki}}
\def\rkj{\rho_{kj}}
\def\rijn{\rho_{12}}
\def\rikn{\rho_{13}}
\def\rjin{\rho_{21}}
\def\rjkn{\rho_{23}}
\def\rkin{\rho_{31}}
\def\rkjn{\rho_{32}}

\def\vab{v_{\!_{AB}}}
\def\vx{\vec{x}}
\def\vxt{\tilde{\vec{x}}}
\def\tT{\tilde{T}}
\def\sn{\sigma_{\!_N}}
\def\xi{x_i}
\def\xj{x_j}
\def\xk{x_k}
\def\xin{x_1}
\def\xjn{x_2}
\def\xkn{x_3}
\def\xc{x_{\!_C}}
\def\xd{x_{\!_D}}
\def\xl{x_{\!_L}}

\def\xlj{x_{\!_{L_j}}}
\def\xlf{x_{\!_{L_1}}}
\def\xln{x_{\!_{L_n}}}

\def\ag{&>&}

\newcommand{\pa}[1]{\left( {#1} \right)}

\newcommand{\pas}[1]{\left[ {#1} \right]}
\newcommand{\Bpa}[1]{\Big(\, {#1} \,\Big)}
\newcommand{\Bpas}[1]{\Big[\, {#1} \,\Big]}
\newcommand{\ave}[1]{\langle {#1} \rangle}

\begin{document}
\begin{frontmatter}
\title{Optional games on cycles and complete graphs}

\author[1,2]{Hyeong-Chai~Jeong\footnote{E-mail:hcj@sejong.ac.kr}}
\author[1]{Seung-Yoon~Oh}
\author[2,3]{Benjamin~Allen}
\author[2,4,5]{Martin~A.~Nowak}

\address[1]{Department of Physics, Sejong University, Gangjingu, Seoul~143-747, KOREA}
\address[2]{Program for Evolutionary Dynamics, Harvard University,Cambridge, MA~20138, USA}
\address[3]{Department of Mathematics, Emmanuel College, Boston,  MA~02115 USA;}
\address[4]{Department of Mathematics, Harvard University,Cambridge, MA~20138, USA}
\address[5]{Department of Organismic and Evolutionary Biology, Harvard University,Cambridge, MA~20138, USA}


\begin{abstract}
We study stochastic evolution of optional games on simple
graphs. There are two strategies, A and B, whose interaction is
described by a general payoff matrix. In addition there are one or
several possibilities to opt out from the game by adopting loner
strategies. Optional games lead to relaxed social dilemmas. Here we
explore the interaction between spatial structure and optional games. 
We find that increasing the number of loner strategies 
(or equivalently increasing mutational bias toward loner strategies) facilitates
evolution of cooperation both in well-mixed and in structured populations.
We derive various limits for weak selection and large population
size. For some cases we derive analytic results for strong selection.  
We also analyze strategy selection numerically for finite
selection intensity and discuss combined effects of optionality
and spatial structure.
\end{abstract}


\end{frontmatter} 

{\bf keywords:} {Evolutionary game theory, Evolutionary graph theory, 
Evolution of cooperation, Spatial games}


\section{Introduction} 
In the typical setting of evolutionary game theory, the individual has
to adopt one of several strategies
\citep{Hofbauer:1988ve,Weibull:1997vv,Friedman:1998uy,Hofbauer:1998wn,Cressman:2003ux,
Nowak:2004vt,Vincent:2005wo,Gokhale:2011hk}.   
For example in a standard cooperative dilemma 
\citep{Hauert:2006kz,Nowak:2012cw,Rand:2013vx,Debarre:2014tq},
the individual can choose between cooperation and
defection. Natural selection tends to oppose cooperation unless a
mechanism for evolution of cooperation is at work 
\citep{Nowak:2006bt}. 
In optional games there is also the possibility not to play the game
\citep{Kitcher:1993tc,Batali:1995uz,Hauert:2002hd,Hauert:2002ua,Szabo:2002cv,DeSilva:2010uz,Rand:2011gy}. 
The individual player has to choose whether to
participate in the game (by cooperating or defecting) or to opt
out. Opting out leads to fixed ``loner's payoff''. This loner's payoff
is forfeited if one decides to play the game. Thus there is a cost for
playing the game. Optional games tend to lead to relaxed social
dilemmas 
\citep{Michor:2002wu,Hauert:2006kz}. 
They have also been used to study the effect of costly punishment (by
peers and institutions) on evolution of cooperation 
\citep{Boyd:1992dt,Nakamaru:2005ug,Hauert:2007js,Sigmund:2007vz,Traulsen:2009cg,Hilbe:2010vs}.
There is also a relationship between optional games and empty places
in spatial settings 
\citep{Nowak:1994vj}.

Here we study the effect of optional games on cycles and on
complete graphs \citep{vanVeelen:2012wl}. 
Cycles and complete graphs are on opposite ends of
the spectrum of spatial structure. Most graphs will lead to an
evolutionary dynamics between these two extremes. 
Evolutionary graph theory
\citep{Lieberman:2005tn,Santos:2005im,Ohtsuki:2006ug,Szabo:2007uy,Fu:2007vo,Fu:2007we,
Santos:2008tg,Perc:2010tr,Perc:2011th,Allen:2013vu,Maciejewski:2014ko,Allen:2014bu}   
is an approach to study the effect of population structure on
evolutionary dynamics 
\citep{Nowak:1992vx,Nakamaru:1997uq,Tarnita:2009vn,Tarnita:2009vna,Nowak:2010ik,Tarnita:2011tf}.
Using stochastic evolutionary dynamics for games in finite populations
\citep{Foster:1990io,Challet:1997wz,Fudenberg:2004ub,Nowak:2004td,Imhof:2006ju,Traulsen:2006uf},
we notice that the
number of different loner strategies has an important effect on
selection between strategies that occur in the game. 
Increasing the number of ways to opt out 
(or, increasing mutational bias toward \citep{Garcia:2012cg} loner strategies) 
in general favors evolution of cooperation.  

Our paper is organized as follows. In Section~2 we give an overview of
the basic model and list our key results. 
In Section 3 we calculate abundance in the low mutation limit.
It is used to investigate the conditions for strategy selection 
in the weak selection limit in Section~4 and 
in the strong selection limit in Section~5. 
We calculate these conditions for optional games
with simplified prisoner's dilemma games in Section~6. 
We then analyze strategy
selection numerically for finite mutation rate as well as finite
selection intensity in low mutation in Section~7.  
In our concluding remarks in Section~8, we summarize and discuss
the implications of our findings.

\section{Model and main results}
\label{s.model}
We consider stochastic evolutionary dynamics of populations on
graphs. In particular, we investigate the condition for one strategy
to be favored over the others in the limit of low mutation and for two
different reproduction processes, birth-death (BD) updating and
death-birth (DB) updating on cycles. We compare the results  with
those for the Moran Process (MP) on the complete graph. 
The fitness of an individual is determined by the payoff
from the non-repeated matrix games with its nearest neighbors.
We use exponential fitness, 
\beqa
 f_r \a= e^{wP_r},
\label{e.fr}
\eeqa 
for the individual at the site $r$, 
where $P_r$ is its accumulated payoff from the games with its neighbors.
The intensity of selection, $w$, is a parameter representing how
strongly the fitness of an individual depends on the its payoff.  

We first study a general matrix game whose payoff matrix 
is given by $A=[\,\aij\,]$, i.e., a game that an individual using 
strategy~$S_i$ receives $\aij$ as a payoff when it plays with an individual 
with strategy~$S_j$. Then we apply our finding to an optional 
prisoner's dilemma game to find a condition for evolution of cooperation.  
 
We calculate abundance (frequencies in the stationary distribution) of
strategies in the low mutation limit, where mutation rate~$u$ goes to zero,
and find the condition that strategy~$S_i$ is more abundant than strategy~$S_j$.
For low mutation, abundance can be written in terms of fixation probabilities
which we obtain in a closed form for general~$w$. 
Although the formal expression of abundance is useful for numerical 
calculation, the complexity of the expression makes it hard for us
to understand the strategy selection mechanism intuitively.

For low intensity of selection ($w\rightarrow 0$),
however, the fixation probability reduces to a linear 
expression in $\aij$ with clear interpretation.
The condition for strategy selection is then given by 
a simple linear inequality in terms of payoff matrix elements.
This is the case even for the large population limit of $N\rightarrow\infty$. 

However, when considering the limits of weak selection ($w \to 0$) and
large population ($N \to \infty$), the condition for strategy selection
depends on the order in which these limits are taken.
We therefore consider two different large population, weak selection
limits: the $wN$ limit and the $Nw$ imit. 
In the $wN$ limit, $w$ goes to zero before $N$ goes to infinity such
that $Nw$ is much smaller than 1. In the $Nw$ limit, 
$N$ goes to infinity before $w$ goes to zero such that 
$Nw$ is much larger than 1. 

\subsection{$wN$ limit}
We first calculate the fixation probability,
$\rik$, which is the probability that a singe $S_i$ takes over the
whole population of the strategy~$S_k$ 
for the $w\rightarrow 0$ limit. It can be written as 
\beqa
 \rik \a= \frac{1}{N} + \dik\, w.
 \label{e.rik}
\eeqa
Here, the ``biased drift'', $\dik$ is defined by
\beqa
\hsg
 \dik \a= \left\{\begin{array}{ll}
      \frac{1}{2}\, \lik  -\frac{1}{2N}\, s_{ik} &\hsg\hsg  \mf \mbox{BD}	\\ 
      \frac{1}{4}\, \lik  -\frac{1}{4N}\, s_{ik} &\hsg\hsg  \mf \mbox{DB}	\\ 
      \frac{1}{4}\, \lik  -\frac{1}{12}\, s_{ik} &\hsg\hsg  \mf \mbox{MP}
 \end{array}
 \right.
\eeqa
with the anti-symmetric term $\lik$ and the symmetric term $s_{ik}$
given by 
\beqa
 \lik \a= \sn \aii + \aik - \aki - \sn \akk \nnn
 s_{ik} \a= \sn \pa{\aii - \aik - \aki + \akk}. 
\eeqa
The structure factor, $\sn$ for the population of the size~$N$, 
is given by 
\beqa
  \sn \a= \left\{\begin{array}{ll}
    1-2/N  & \mf \mbox{BD\ \& \ MP}	\\ 
    3-8/N  & \mf \mbox{DB}.
 \end{array}
 \right.
\label{e.sn}
\eeqa

Using fixation probabilities of Eq.~(\ref{e.rik}), we then calculate
abundance in the low mutation limit and show that strategy~$S_i$ is more
abundant than strategy~$S_j$ when
\beqa
 \sum_{k}\, \lik \ag \sum_{k}\, \ljk
\eeqa
as previously known \citep{Nowak:2010ik,Ohtsuki:2006gp}.
The fixation probability obtained for a general $2\times 2$ matrix
game is also applied to calculate abundance of cooperator and
defectors in optional prisoner's dilemma game\citep{Szabo:2002gd} with $(n+2)$ strategies,  
cooperator (C), defector (D) and $n$ different types of loners, $L_1,\cdots,L_n$. 
The payoff matrix is given by  
\beqa
\hsp{-10\tm}
\begin{array}{cl}
 \hsc    & \begin{array}{ccccc}
     \hsb   C  &\hsa\ \ D &\hsa\ \ L_1 &\ \cdots &\ L_n 
       \end{array} \ \	\\
\begin{array}{c}
      C      \\
      D      \\
      L_1      \\
      \vdots   \\
      L_n
\end{array}
   & \left( \begin{array}{ccccc}
\hsa    R   & \hsb  S     &\hsb  g    &\hsa \cdots  &\hsb  g     \\
\hsa    T   & \hsb  P     &\hsb  g    &\hsa \cdots  &\hsb  g     \\
\hsa    g   & \hsb  g     &\hsb  g    &\hsa \cdots  &\hsb  g     \\
\hsa \vdots & \hsb \vdots &\hsb\vdots &\hsa \ddots  &\hsb\vdots  \\
\hsa    g   & \hsb  g     &\hsb  g    &\hsa \cdots  &\hsb  g     \\
  \end{array} \right).  
\end{array}
\label{e.CDLn}
\eeqa
When two cooperators meet, both get payoff $R$. 
When two defectors meet, they get payoff $P$. 
If a cooperator meets a defector, the defector gets the payoff $T$ 
while the cooperator get the payoff $S$.
Loners get payoff $g$ always. Cooperators or defectors also get
payoff $g$ when they meet a loner. 
Since the $n$ different types of loners have the same payoff
structure, this system is equivalent to 
to the population with three strategies, $C$, $D$, and a single
type of loners, $L$ if the mutation rate toward $L$ (from $C$ or $D$)
is $n$ times larger than the other way. 

In the limit of $w$ goes to zero, we find that the condition for $\xc>\xd$
is given as 
\beqa
 \sn(n)\, R + S &>&  T + \sn(n)\, P,
\label{e.rstpgw1}
\eeqa
where
\beqa \hsg\hsg\hsg
 \sn(n) \sa= \left\{\begin{array}{lll}
  \pa{1 + \frac{1}{2}\,n}\pa{1-\frac{2}{N}} \hsg  & \hsg \mf \mbox{BD\ \&  MP}	\\ 
  \pa{1 + \frac{1}{2}\,n}\pa{3-\frac{8}{N}} \hsg  & \hsg \mf \mbox{DB}.
 \end{array}
 \right.
\eeqa
As long as $w$ goes to zero first ($Nw\ll 1$),
inequality~(\ref{e.rstpgw1}) is valid 
even in the large population limit of $N\rightarrow\infty$,
where the structure factor, $\sn(n)$ becomes
\beqa \hsg\hsg
 \sigma(n) \sa= \left\{\begin{array}{lll}
   1+\frac{1}{2}\,n  & \mf \mbox{BD\ \& MP}	\\ 
   3+\frac{3}{2}\,n  & \mf \mbox{DB}.
 \end{array}
 \right.
\eeqa

If we do not allow any loner type, then $n=0$ and $\sn(n)$ becomes
$\sn$ of Eq.~(\ref{e.sn}) as expected, and cooperators are more abundant
than defectors if and only if $\rcd > \rdc$.
On the other hand, when the number of loner types, $n$, goes to infinity, 
$\sigma$ becomes infinity and social dilemmas are completely resolved. Cooperators 
are more abundant than defectors whenever $R>P$.  

\subsection{Nw limit}

We still consider the low selection intensity limit ($w\rightarrow 0$) 
but we take the large population limit first such that $Nw$ is much larger
than 1. In this case, we can calculate the fixation
probability analytically only for BD and DB. 
Fixation of~$S_i$ (invading strategy~$S_k$) is possible only when 
$\lik$ is positive where
\beqa
 \lik \a= \sigma \aii + \aik - \aki - \sigma \akk.
 \label{e.lik}
\eeqa
The structure factor for infinite population, $\sigma$ in Eq.~(\ref{e.lik})
is 1 for BD and 3 for DB.
When $\lik$ is positive, fixation probability, $\rik$ is proportional to 
$\lik$ and given by 
\beqa
 \rik \a= \left\{\begin{array}{ll}
          \, \lik \,\Theta(\lik) &    \mf \mbox{BD}	\\ 
          \half \lik\, \Theta(\lik) &    \mf \mbox{DB},
 \end{array}
 \right.
\label{e.rikNw}
\eeqa
where $\Theta(x)$ is the Heaviside step function.
 
We calculate abundance for the low mutation limit
using fixation probabilities given by Eq.~(\ref{e.rikNw}) 
for a general 3 strategy game and 
find conditions for the abundance $x_i$ of strategy $S_i$ to be larger
than the abundance $x_j$ of strategy $S_j$. 
Here $i$, $j$, and $k$ are the indices representing
three distinct strategies, $S_i$, $S_j$, and $S_k$. 
If both $\lij$ and $\lik$ are positive, 
$S_j$ and~$S_k$ cannot invade~$S_i$ and we have
$\xi=1$ and $\xj=\xk=0$, i.e., $\xi>\xj$ always. 
By the same token, $\xi$ cannot be larger than 
$\xj$ when both $\lji$ and $\ljk$ are positive.
If $\lki$ and $\lkj$ are positive, both 
$\xi$ and $\xj$ are zero. The only 
non-trivial case is when three strategies, 
show rock-paper-scissors-like characteristics
in terms of $\lij$. 
For the $\lij>0$ case (with $\ljk>0$ and $\lki>0$),
strategy~$S_i$ is more abundant than strategy~$S_j$
when $\lij > \lki$. 
For the $\lji>0$ case (with $\lik>0$ and $\lkj>0$),
strategy~$S_i$ is more abundant than strategy~$S_j$ 
when $\lji < \lkj$. 

The analysis for three strategy game can be applied 
to optional prisoner's game with $n$ types of loners
whose payoff matrix is given by Eq.~(\ref{e.CDLn}). 
The condition for $\xc>\xd$ can be  still written as a linear inequality
but the coefficients of the linear inequality
depend on the signs of~$R-P$, $R-g$, and $P-g$. 
For simplicity, we first assume that $R>P$ without loss of generality.  
Then, when $P>g$, the condition for $\xc> \xd$ becomes
\beqa
\begin{array}{llll}
\ R + S & > & T +\ P & \mf \mbox{BD}	\\ 
 3R + S & > & T + 3P & \mf \mbox{DB}.
\end{array}
\label{ie.rstp.liss1}
\eeqa
For the other case of $P<g$, the condition for $\xc> \xd$ becomes
\beqa
\begin{array}{llll}
\ R + S & > & T +\ P +\ n(P-g) & \mf \mbox{BD}	\\ 
 3R + S & > & T + 3P + 3n(P-g) & \mf \mbox{DB}.
\end{array}
\label{ie.rstp.liss2}
\eeqa

For high intensity of selection ($w\gg 1$), strategy selection
strongly depends on the number of loner strategies, $n$. 
If $n$ is larger than 1, cooperators are more abundant than defectors
as long as $g>P$. On the other hands, for $n=1$, the condition 
for $\xc>\xd$ depends on the reproduction processes. For $n=1$, we obtain
the condition only for the ``simplified'' prisoner's dilemma game
(``donation game'') in
which the payoffs are described in terms of the benefit, $b$ and the
cost, $c$ of cooperation, $R=b-c$, $S=-c$, $T=b$, $P=0$.
For BD, cooperators are always less abundant than defectors as long as
$g<b$. For DB and MP, $\xc$ is larger than $\xd$ if
\beqa
\begin{array}{llll}
 c  & < & b/2  & \mf \mbox{DB}	\\ 
 c  & < & g    & \mf \mbox{MP}.
\end{array}
\eeqa

\subsection{Numerical analysis}

Our analytic results are obtained in the two extreme limits of selection
strength ($w\to 0$ and $w\to\infty$) in the zero mutation limit. 
For finite values of~$w$ (with low mutation rate),
we solve conditions for $\xc>\xd$ numerically, using calculated abundance
from fixation probabilities. For finite mutation rates, 
we perform a series of Monte Carlo simulations and 
measure abundance to obtain the condition for strategy selection. 

In particular, we consider a simplified prisoner's dilemma game 
with one type of loners ($n=1$) in which  
the analytic conditions for $\xc>\xd$ [inequalities~(\ref{e.rstpgw1}),
(\ref{ie.rstp.liss1}) and  (\ref{ie.rstp.liss2})] become
\beqa
 \begin{array}{llll}
  c &<&\frac{N-6}{5N-6} \, b     & \mf \mbox{BD\ \&  MP}	\\ 
  c &<&\frac{7N-24}{11N-24} \, b & \mf \mbox{DB},
 \end{array}
\label{e.cri.ws}
\eeqa
for the $wN$ limit, and 
\beqa
\begin{array}{llll}
  g & > & 2c & \mf \mbox{BD}	\\ 
  g & > & \frac{4}{3}\pa{c-b/2} & \mf \mbox{DB},
\end{array}
\label{e.cri.ss}
\eeqa
for the $Nw$ limit.

We first confirm these conditions numerically with 
a finite but small $w$ in the low mutation limit. 
Abundance of each strategy is calculated 
for $Nw=0.01$ and $Nw=100$ (with $N=10^4$).
We find more cooperators than defectors when 
inequality~(\ref{e.cri.ws}) is satisfied for $Nw=0.01$ 
and inequality~(\ref{e.cri.ss}) for $Nw=100$. 
When $Nw$ is much smaller than 1, cooperators in BD and those in MP
are more abundant than defectors in the same region in the parameter space
as inequality~(\ref{e.cri.ws}) predicts. However, they are different
for general $Nw$. When $Nw$ is much larger than  1, cooperators are
less abundant than defectors always for BD but we find more cooperators than
defectors when $g>c$ for MP.

For finite mutation rate, we investigate abundance by Monte Carlo simulation.
We start from a random arrangement of three strategies on a cycle (BD and DB)
or a complete graph (MP) with $N=50$ sites.  
Population evolves with BD, DB, or MP updating processes
with the mutation rate, $u=0.0002$. 
We monitor the time evolution of the average frequencies and 
see if the population evolves to a steady state
in which average frequency remains constant. We measure
abundance, the frequency average in the steady state,
and find that abundance in our simulations agrees quite well
with calculations in the low mutation limits using fixation probabilities. 

\section{Derivation of general expressions for fixation probability and abundance} 

We now begin our derivation of the results presented above. 
We begin by obtaining general expressions for fixation probability and
abundance that are valid for any population size and selection
intensity. These expressions are obtained first for a general 
$3\times 3$ matrix game, and then for the optional prisoners' dilemma game. 

When there are mutations, the population will not evolve to an absorbing 
state of one kind. Yet, in many cases, it is expected for them to evolve to a
steady state in which the frequency of each type (in a sufficiently large
population) stays constant. We use the term ``abundance'' for
frequency in the steady state. For a small population, frequencies may
oscillate with time through mutation-fixation cycles, especially when the
mutation rate is very small. In this case, abundance is defined as the
time average of frequencies over fixation cycles. 

In this section, we consider abundance in the low mutation limit, in
which the mutation rate $u$ goes to zero. We imagine an invasion of a
mutant in the mono-strategy population and we ignore the possibility of
further mutation during the fixation sweep. In this low mutation limit,
abundance can be expressed in terms of fixation probabilities.  
We first calculate fixation probabilities for general selection 
intensity~$w$ and present them in a closed form for BD and DB. 
Then, we present abundance in terms of fixation probabilities. 

\subsection{Fixation probability}
\label{s.fp}
We consider the fixation probability of~A (invading a population that
consists of~B) for a general 2x2 matrix game with the payoff matrix,
\[
\begin{array}{cl}
     & \begin{array}{cc}
        \hsa\ \ A  & \hsa \ \ B \hsb 
       \end{array} \ \	\\
\begin{array}{c}
     A      \\
     B
\end{array}
   & \left( \begin{array}{cc}
   a   & \hsb  b\\
   c   & \hsb  d  
  \end{array} \right).  
\end{array}
\]
In general, the fixation probability of~$A$ is given by
\beqa
 \rab \a= \pas{1+\sum_{m=1}^{N-1} \prod_{\na=1}^m \frac{T_\na^-}{T_\na^+}}^{-1}
\label{e.fix}
\eeqa
where $T_\na^\pm$ is the probability that the number of~A
becomes $\na\pm 1$ from $\na$~\citep{Nowak:2006vy}.
When new offspring appear in nearest neighbor sites, as they do for
BD and DB, only one connected cluster of invaders
can form on a cycle and $T_\na^\pm$ can be easily calculated. 
In fact, with exponential fitness, $\rab$ is given in a closed form.

For BD, the fixation probability can be written in the form of 
\beqa 
  \rab \a= \frac{f}{g\ + \ h\, y^N},
\label{e.fixBD}
\eeqa
with 
\beqa 
\hsg\hsg   f \sa= e^{w(a+b)}-e^{w(c+d)}\nnn
\hsg\hsg  g \sa= e^{w(a+b)}-e^{w(c+d)}+e^{w(a-b+c+d)} \nnn
\hsg\hsg   h \sa= e^{w(2a-c-2d)}\pas{e^{w(a+b+c)}-e^{w(a+b+d)}-e^{w(2c+d)}} \nnn
\hsg\hsg  y \sa= e^{w(c+d-a-b)},
\label{e.fixBDf}
\eeqa       
when $a+b\not= c+d$. Note that both denominator and numerator of the
right hand side of Eq.~(\ref{e.fixBD}) are zero when $a+b=c+d$.
For this singular case, $\rab$ can be directly calculated from 
Eq.~(\ref{e.fix}) and is given by
\beqa \hsg\hsg\hsg
  \rab \sa= \frac{1}{1\ + e^{2w(b-c)} - 2\,e^{w(a-b)} + N\, e^{w(a-b)}}.
\label{e.fixBDs}
\eeqa
In the limit of $a+b \rightarrow c+d$, 
Eq.~(\ref{e.fixBD}) [with Eq.~(\ref{e.fixBDf})]
becomes identical to Eq.~(\ref{e.fixBDs}).
Hence, we can write the fixation probability of~A for BD on a cycle 
as Eq.~(\ref{e.fixBD}) for general case if it is understood as the
limiting value when both denominator and numerator becomes zero. 

For DB, the fixation probability can be also written in the form of Eq.~(\ref{e.fixBD})
but now with 
\beqa
\hsp{-5\tm} f \a=  e^{w(3a+b)}-e^{w(c+3d)} \hsd \nnn
\hsp{-5\tm} g \a= \pa{e^{w(3a+b)}-e^{w(c+3d)}}\frac{3+e^{2w(d-b)}}{2}\nnn
\hsp{-5\tm} & & +\, \frac{e^{w(c+d)}\pa{e^{2wb}+e^{2wd}}
              \pa{e^{w(a+b)}+e^{2wd}}\pa{e^{2wa}+e^{w(c+d)}}}
                 {2e^{2wb}\pa{e^{w(a+b)}+e^{w(c+d)}}}\nnn
\hsp{-5\tm} h \a=\pas{e^{w(3a+b)}\pa{e^{2wb}+e^{2wd}}\pa{e^{2wa}+e^{w(c+d)}}^4} \nnn
\hsp{-5\tm}&&\times\pas{\frac{\pa{e^{2wa}+3e^{2wc}}\pa{e^{w(c+3d)}-e^{w(3a+b)}}}
             {2e^{3w(c+d)}\pa{e^{w(a+b)}+e^{2wd}}^4\pa{e^{2wa}+e^{2wc}}}}\nnn
\hsp{-5\tm}&& -\, \frac{e^{2w(2a+b)}\pa{e^{2wb}+e^{2wd}}\pa{e^{2wa}+e^{w(c+d)}}^5}
           {2e^{3w(c+d)}\pa{e^{w(a+b)}+e^{2wd}}^3\pa{e^{w(a+b)}+e^{w(c+d)}}}\nnn
\hsp{-5\tm} y \a= \frac{e^{-w(3a+b-c-3d)} +  e^{w(c+d-2a)}}
             {1\ +\ e^{w(c+d-2a)}},
\label{e.fixDBf}
\eeqa       
when $3a+b\not= c+3d$. 
We can also show that 
Eq.~(\ref{e.fixBD}) [with Eq.~(\ref{e.fixDBf})]
becomes the fixation probability for $3a+b=c+3d$
if we take the limit of $3a+b \rightarrow c+3d$.

For MP, the fixation probability given by Eq.~(\ref{e.fix}) cannot be
written in a closed form in general but reduces~\citep{Traulsen:2008ee} to 
\beqa \hsbb 
   \rab \sa= \pa{\sum_{m=1}^{N-1}
     e^{-w\pas{\frac{(a-b-c+d)}{2(N-1)}\,m(m+1)
         -\frac{(a-bN+dN-d)}{N-1}\,m}}}^{-1}.
\label{e.fixMPa}
\eeqa
For $a+d=b+c$, the summation in Eq.~(\ref{e.fixMPa}) can be calculated exactly
and we have 
\beqa \hsbb
  \rab \sa= \frac{e^{w(a-bN+dN-d)/(N-1)}-1}{e^{wN(a-bN+dN-d)/(N-1)}-1}.
\label{e.fixMPs}
\eeqa
For $a+d \not= b+c$, the summation can be approximated by an integral~\citep{Traulsen:2008ee}
and we have 
\beqa \hsbb
  \rab \saa=\frac{{\rm erf}\Bpa{\sqrt{w\over u}\,[u+v]}-{\rm erf}\Bpa{\sqrt{w\over u}\,v}}
                   {{\rm erf}\Bpa{\sqrt{w\over u}\,[uN+v]}-{\rm erf}\Bpa{\sqrt{w\over u}\,v}}.
\label{e.fixMPi}
\eeqa
Here, $u=(a-b-c+d)/(2N-2)$, $v=(-a+bN-dN+d)/(2N-2)$ and
${\rm erf}(x) = \frac{2}{\sqrt{\pi}}\, \int_0^x e^{-y^2}\, dy$
is the error function.  
The summation in Eq.~(\ref{e.fixMPa}) can be also calculated exactly
for the $wN$ limit (see Section~\ref{s.wN}) where the exponential term
can be linearized. 

\subsection{Abundance in the low mutation limit} 
Let $\xi$ be the abundance of strategy $S_i$, whose payoff matrix is given by 
\beqa
\begin{array}{cl}
 \hsc    & \begin{array}{ccc}
     \hsb   S_1  &\hsb\ \ S_2 &\hsb\ \ S_3 
       \end{array} \ \	\\
\begin{array}{c}
      S_1      \\
      S_2      \\
      S_3 \end{array}
   & \left( \begin{array}{ccc}
\hsa    a_{11}   & \hsb  a_{12}     &\hsb  a_{13} \\
\hsa    a_{21}   & \hsb  a_{22}     &\hsb  a_{23} \\
\hsa    a_{31}   & \hsb  a_{32}     &\hsb  a_{33} 
  \end{array} \right).  
\end{array}
\label{e.pm3}
\eeqa
Then, in the low mutation limit, we expect the abundance vector,
$\vx = (\xin,\xjn,\xkn)$ can be written as 
\beqa
 \vx &=& \vx\, T,
\eeqa
with the transfer matrix
\beqan 
 T &=& \begin{pmatrix}
    1-\rjin-\rkin & \rjin         & \rkin \\
    \rijn         & 1-\rijn-\rkjn & \rkjn \\
    \rikn         & \rjkn         & 1-\rikn-\rjkn 
    \end{pmatrix}.
\eeqan
Here, $\rij$ is the fixation probability of strategy~$S_i$ (invading the
population of strategy~$S_j$).
A (unnormalized) left eigen-vector of $T$ with the unit eigen value,
$\vx^u=(\xin^u, \xjn^u, \xkn^u)$ is given by
\beqa
\xin^u \a= \rijn \rikn\, +\, \rikn \rkjn  + \rijn \rjkn \nnn
\xjn^u \a= \rjkn \rjin\, +\, \rjin \rikn  + \rjkn \rkin \nnn
\xkn^u \a= \rkin \rkjn\, +\, \rkjn \rjin  + \rkin \rijn. 
\label{e.eigen}
\eeqa
Once we calculate all fixation probabilities $\rij$,
the steady state frequencies, $\xi$ can be obtained 
by normalizing~$\xi^u$;
\beqa
 \xi \a=  \xi^u\, /\, \sum_{j} \xj^u.
\label{e.steady_x}
\eeqa

\subsection{Optional prisoner's dilemma game}
The fixation probabilities obtained in Section~\ref{s.fp} can be
used to calculate abundance of cooperators and defectors
in optional prisoner's dilemma game. Here, we consider the
game with $(n+2)$ strategies, 
cooperator (C), defector (D) and $n$ different loners,
$L_1,\cdots,L_n$ whose payoff matrix is given by 
Eq.~(\ref{e.CDLn}).
We introduce $n$ different types of loners 
to investigate how the condition for the emergence of cooperation varies
with the number of loner types, $n$.  

Let $\xc$, $\xd$, and $\xlj$ be the abundance of $C$, $D$, and $L_j$, respectively.
Then, for low mutation, 
the abundance vector $\vxt = (\xc,\xd,\xlf,\cdots,\xln)$ can be written
as  
\beqa
 \vxt \a= \vxt\, \tilde{T}
\label{e.vxt}
\eeqa
with 
\beqa 
 \tT  \a= \begin{pmatrix}
     \tT_{CC}  \hsa & \rdc       \hsa & \rlfc        & \cdots     & \rlnc \\
     \rcd     \hsa & \tT_{DD}    \hsa & \rlfd        & \cdots     & \rlnd \\
     \rclf    \hsa & \rdlf       \hsa & \tT_{L_1L_1}  & \cdots     & \rlnlf \\
\hsa \vdots   \hsa & \hsb \vdots \hsa &\hsb\vdots    &\hsa \ddots &\hsb\vdots  \\
     \rcln    \hsa & \rdln       \hsa & \rlfln       & \cdots     & \tT_{L_nL_n}
     \end{pmatrix}.
\label{e.Tt}
\eeqa
As before, $\rij$ is the fixation probability that an~$S_i$ takes over 
the population of~$S_j$ 
and $\tT_{ii} = 1-\sum_{j\not=i} \rij$
with the convention that strategy~1 is C, 
strategy~2 is D, and 
strategy~$S_i$ is $L_{i-2}$ for $i>2$. 
Since the payoffs of the games involving loners are independent
of the loner type, so are the fixation probabilities
involving~$L_j$. By denoting $\rxlj$ by $\rxl$,
Eqs.~(\ref{e.vxt}) and~(\ref{e.Tt}) can be rewritten 
in terms of the total frequency of loners $\xl = \sum_j \xlj$ as
\beqa
 \vx \a= \vx\, T
\label{e.vxs}
\eeqa
with  $\vx = (\xc,\xd,\xl)$, where
\beqa 
 T \a= \begin{pmatrix}
    1-\rdc-n\rlc\hsg\hsg & \rdc         & n \rlc \\
    \rcd         &\hsg\hsg 1-\rcd-n\rld & n \rld \\
    \rcl         & \rdl         &\hsg\hsg 1-\rcl-\rdl 
    \end{pmatrix}.
\label{e.TLn}
\eeqa
The evolution dynamics of Eq.~(\ref{e.vxs}) with the transfer matrix, 
$T$ of Eq.~(\ref{e.TLn}) can be interpreted as biasing the mutation 
rate toward loner strategies. The mutation rate toward $L$ (from $C$ or $D$)
is $n$ times larger than the other way. 

The abundance vector of three strategies, C, D, and L, is proportional to 
the left eigen-vectors of $T$ with the unit eigen value,
$\vx^u=(\xc^u, \xd^u, \xl^u)$, given by
\beqa
\xc^u \a=\ \  \rcd \rcl\, +n\, \rcl \rld  +\ \ \rcd \rdl  \nnn
\xd^u \a=\ \  \rdl \rdc\, +\ \ \rdc \rcl  +n\, \rdl \rlc  \nnn
\xl^u \a= n^2 \rlc \rld   +n\, \rld \rdc  +n\, \rlc \rcd. 
\label{e.eigenCDL}
\eeqa
Here $\rcd$, $\rcl$, $\ldots$ are fixation probabilities 
between three strategies with payoff matrix,
\beqa
\begin{array}{cl}
     & \begin{array}{ccc}
       \hsb C  & \hsa\ \ D & \hsb L 
       \end{array} \ \	\\
\begin{array}{c}
     C\hsg \\
     D\hsg \\
     L\hsg 
\end{array}
   & \left( \begin{array}{ccc}
   R   &\hsb S & \hsb\ g \\
   T   &\hsb P & \hsb\ g \\ 
   g   &\hsb g & \hsb\ g  
  \end{array} \right).  
\end{array}
\label{e.pmCDL}
\eeqa

\section{Analysis of the $wN$ limit} 
 \label{s.wN}

We now consider the results of Section 3 under the $wN$ limit.
This limit is obtained by taking the $w \to 0$ limit for fixed $N$, and
then taking the $N \to \infty$ limit of the result. 
We calculate abundance in terms of fixation probabilities in the $wN$
limit and analyze the condition for the cooperators are more abundant
than defectors. 

\subsection{Fixation probability}  
As $w$ goes to zero, the fixation probability for BD, 
Eq.~(\ref{e.fixBD}) [with Eq.~(\ref{e.fixBDf})] becomes
\beqa 
 \rab \a= \frac{1}{N}\, +\, \frac{w}{2N^2}\Bpas{\pa{N^2-3N+2}a + \pa{N^2+N-2}b} \nnn  
        & & \hsb -\, \frac{w}{2N^2}\Bpas{\pa{N^2-N+2}c  + \pa{N^2-N-2}d} \nnn
      \a= \frac{1}{N} + \frac{w}{2}\Bpas{\pa{\sn a + b - c -\sn d}  -  \frac{\sn}{N}\pa{a-b-c+d}},
\label{e.rab.bd} 
\eeqa
where $\sn= 1-2/N$. In the second line, we divide
the $w$~dependent parts as the sum of the anti-symmetric term and
the symmetric term under exchange of~A and~B. 
The symmetric term contributes equally to both $\rab$ and $\rba$
and is irrelevant to determine abundance.
For DB, the fixation probability according to Eq.~(\ref{e.fixBD}) 
[with Eq.~(\ref{e.fixDBf})] becomes
\beqa 
 \rab \a= \frac{1}{N}\, + 
            \frac{w}{4N^2}\Bpas{\pa{3N^2- 11N+ 8}a +\pa{N^2+3N-8}b}  \nnn
      & &\hsb - \frac{w}{4N^2}\Bpas{\pa{N^2-3N+8}c  +\pa{3N^2-5N-8}d} \nnn
      \a= \frac{1}{N} + \frac{w}{4}
        \Bpas{\pa{\sn a+ b- c - \sn d} - \frac{\sn}{N}\pa{a-b-c+d}},
\label{e.rab.db} 
\eeqa
where $\sn= 3-8/N$.
For MP, the fixation probability cannot be expressed in a closed form for
general~$w$. However, when $w$ goes to zero, it can be calculated using
Eq.~(\ref{e.fixMPa}), and is given by 
\beqa
 \rab \sa= \frac{1}{N} + \frac{w}{6N}\Bpas{(N-2)a +(2N-1)b} 
            - \frac{w}{6N}\Bpas{(N+1)c + (2N-4)d} \nnn
     \sa= \frac{1}{N} + \frac{w}{4}\Bpas{\sn a +b -c -\sn d} 
            - \frac{w}{4}\Bpas{\frac{\sn}{3}\pa{a - b - c + d}},
\label{e.rab.mp} 
\eeqa
where $\sn=1-2/N$. 

The fixation probabilities for the three processes, 
as given by Eqs.~(\ref{e.rab.bd}-\ref{e.rab.mp}), 
can be expressed as
\beqa
 \rab \sa= \frac{1}{N}  + w \theta_a  \Bpas{\sn a + b - c - \sn d} 
          - w \theta_s  \Bpas{\sn \pa{a - b - c + d}}, 
\label{e.fix.w}
\eeqa
with $\sn$, $\theta_a$, and $\theta_s$ given by the following table. 
\beqa 
 \begin{array}{||c||c|c|c||} \hline \hline
            & \sn              & \hsa \theta_a\hsa &\hsa \theta_s\hsa \\ \hline \hline
  \mbox{BD} & 1-\frac{2}{N}    & \frac{1}{2}       &\frac{1}{2N}     \\ \hline
  \mbox{DB} & 3-\frac{8}{N}    & \frac{1}{4}       &\frac{1}{4N}     \\ \hline
  \mbox{MP} & 1-\frac{2}{N}    & \frac{1}{4}       &\frac{1}{12}     \\ \hline \hline
 \end{array}
\label{e.sntheta}
\eeqa
We would like to emphasize that the difference between $\rab$ and $\rba$ 
comes form the anti-symmetric term. In other words, strategy selection is
determined by the sign of $\sn a + b - c + \sn d$. This value 
is identical for BD on cycle and MP.
The coefficient of the anti-symmetric term, $\theta_a$ for BD and MP would have been the same if we
had normalized the accumulated payoff such that 
an individual in a population of mono-strategy  
has the same fitness both for BD and MP. 
For MP, each individual plays games with $N-1$ neighbors 
while an individual on a cycle has two neighbors. To have the same
effective payoff with individual on a cycle,  we need to normalize the
accumulated payoff for MP by multiplying $2/(N-1)$. However, for MP, we
use $P_r$ in Eq.~(\ref{e.fr}) as the average payoff which is the 
accumulated payoff divide by $N-1$, following the established
convention~\citep{Nowak:2006vy}.  
Hence, the results for MP using intensity of selection, $w$ should
be compared with those with half of the intensity, $w/2$ for BD
and DB.
We also note that the symmetric terms are of order $w/N$ 
for BD and DB on cycles while it is of order $w$ for MP.

Fixation probability in the $wN$ limit is obtained by taking
$N\rightarrow \infty$ limit of Eq.~(\ref{e.fix.w}) and Eq.~(\ref{e.sntheta});
\beqa 
 \rab \a= \left\{\begin{array}{ll}
   \frac{1}{N}\pas{1 + \frac{Nw}{2}\pa{a+b-c-d}}   &\hsbb  \mf \mbox{BD}	\\ 
   \frac{1}{N}\pas{1 + \frac{Nw}{4}\pa{3a+b-c-3d}} &\hsbb \mf \mbox{DB}	\\    
   \frac{1}{N}\pas{1 + \frac{Nw}{6}\pa{a+2b-c-2d}} &\hsbb  \mf \mbox{MP}.
 \end{array}
 \right. 
\label{e.Nwfix}
\eeqa

These results can be understood by considering fixation process as a
(biased) random walk on a one-dimensional lattice.   
Let $T_\na^\pm$ be the probability that the number of A to be $\na\pm 1$
from $\na$ as introduced in Eq.~(\ref{e.fix}). 
Then, without a mutation, we have $T^-_N = T^+_0 = 0$.
Hence, there are two absorbing states, the all~B state at $\na=0$ and
the all~A state at $\na=N$.  
Now, $\rab$ can be interpreted as
the probability that the random walker reaches
the $\na=N$ state starting from the $\na=1$ state. For large $N$, the
master equation describing population dynamics can be approximated by
a Fokker-Plank equation with (biased) drift, $v_\na$, and the
(stochastic) diffusion,  $d_\na$, which are approximately given by  
$v_\na \approx (T_\na^+ - T_\na^-)$ and 
$d_\na \approx \pa{T_\na^+ + T_\na^-}/N$~\citep{Traulsen:2006uf}.
For small $w$, drift velocity is proportional 
to~$w$, and the relative contribution of the diffusion term, 
$d_\na/v_\na$ is asymptotically given by
$\frac{d_\na}{v_\na} \sim \frac{1}{Nw}$.
For weak selection ($Nw\ll 1$), where $d_\na/v_\na$ is large, the fixation probability is 
mainly determined by the (stochastic) diffusion term, $1/N$
and can be written as
\beqa
  \rab = \frac{1}{N} + \vab.
 \label{e.rabrw}
\eeqa 
The perturbation term, $\vab$ is the (weighted) average 
drift velocity over $\na=1$ to $\na=N-1$ state and is given by 
\beqa
 \vab \a= \ave{v_\na}
      \b= \sum_\na \,\pna\ (T_\na^+ - T_\na^-),
 \label{e.vice}
\eeqa
where $\pna$ is the frequency of visits to the state $\na$ (the expected sojourn time at $\na$). 
When $w$ is small, the difference between $T_\na^+$ and $T_\na^-$ is
also small and ``walkers'' can diffuse around state $\na$ easily.
Then we can treat $x=\na/N$ as a continuous variable, especially
when $N$ is large. Hence, for small~$w$ and large~$N$, $\phi$
satisfies the diffusion equation in one-dimension,
\beqa
\frac{d^2\phi}{dx} \a= 0, 
\eeqa
whose solution is given by 
\beqa
 \pna \a= c_1 + c_2\, \frac{\na}{N} \nnn
      \a= \frac{2}{N(N-1)}\, \pas{(N-1)-(N-2)\frac{\na}{N}} \nnn
      \aa= \frac{2}{N}\, \pa{1-\frac{\na}{N}},
\label{e.phi}
\eeqa
for $\na=1,\cdots,N-1$. 
Here, two constants $c_1$ and $c_2$ have been determined by the 
boundary conditions, 
$\phi_{_{\!N}} = \frac{1}{N-1} \phi_0$
(for neutral drift of $w=0$,
$\frac{\phi_{_{\!N}}}{\phi_0} = \frac{1/N}{1-1/N} = \frac{1}{N-1}$) 
and the normalization, $\sum \pna = 1$.

Since $T_\na^{\pm}=T^\pm$ is independent of $\na$ for 
almost every~$\na$, for BD (except $\na=1$ and $\na=N-1$) and 
DB (except $\na=1$, 2, $N-2$, and $N-1$) on cycles, 
$\vab$ can be treated as a constant for large~$N$. 
By considering the motion of the domain boundary 
between A and B blocks, we obtain
\beqa \hsbb
  \vab \sa= \ave{T^+ - T^-} \nnn \hsbb
       \sa=  \left\{\begin{array}{ll}
	\frac{w}{2}\pa{a+b-c-d}  &\hsg \mf  \mbox{BD}	\\ 
        \frac{w}{4}\pa{3a+b-c-3d}&\hsg \mf  \mbox{DB}.
	\end{array}
	\right.
 \label{e.vabrwbd}
\eeqa
For MP, $T_\na^\pm$ depends $\na$ but
$\vab$ can be also easily calculated from $\pna$
of Eq.~(\ref{e.phi}). During the fixation sweep, the average number of~$A$
in the population is $\ave{\na} = \sum\, \na\, \pna \approx N/3$. 
In the $wN$ limit, we have
\beqa \hsbb
 \vab \hsg \aa= \hsg \frac{w}{2} \sum_\na \pna \pas{(a-c)\, \na + (c-d) (N-\na)}\nnn
      \sa= \frac{Nw}{6} \pas{a-c + 2(b-d)}.
 \label{e.vabrwmp}
\eeqa            
Inserting $\vab$ given by Eq.~(\ref{e.vabrwbd}) or~(\ref{e.vabrwmp}),       
into Eq.~(\ref{e.rabrw}), we recover Eq.~(\ref{e.Nwfix}).

\subsection{Strategy selection} 
\label{ss.se}
Here, we consider the condition for the strategy~$S_i$ is
more abundant than the strategy~$S_j$, i.e., $\xi>\xj$. 
We can write the formal expression for 
the condition $\xi>\xj$ for the general selection 
strength and population size using 
Eqs.~(\ref{e.eigen}) and~(\ref{e.fixBD}).
Although the formal expression may be useful to analyze abundances of
strategies numerically, it provides little analytic 
intuition due to the complexity of the expression. 
Hence, here, we solve the inequalities analytically 
for low intensity of selection ($w\rightarrow 0$). For finite
intensity of selection, we find the condition for $\xi>\xj$
numerically in Section~\ref{s.na}.

When $wN$ is much smaller than 1, from Eq.~(\ref{e.fix.w}), the
fixation probability, $\rij$ is written as  
\beqa
 \rij  \a= \frac{1}{N}\pas{1+w \dij}
\label{e.gfix.w}
\eeqa
with
\beqa \hsg
 d_{ij} \sa= \theta_a \pa{\, \sn\aii + \aij - \aji - \sn\ajj\,}
          - \theta_s \sn \pa{\,\aii-\aij-\aji +\ajj\,}.
\label{e.dij}
\eeqa 

Since abundance $x_1$ of strategy~$S_1$ is proportional to $x_1^u$ of
Eq.~(\ref{e.eigen}), we can write, 
\beqa 
 x_1 \sap= \pa{1+w\dijn}\pa{1+w\dikn} + \pa{1+w\dikn}\pa{1+w\dkjn} 
       + \pa{1+w\dijn}\pa{1+w\djkn}\nnn \hsbb
     \saa= 3 + w\pas{2(\dijn+\dikn)+\dkjn+\djkn}  \nnn\hsbb
     \sa= 3\!+\!w\pas{(\dijn\!-\!\djin)\! +\! (\dikn\!-\!\dkin)} 
        + w\pas{(\dijn\!+\!\djin)+(\dikn\!+\!\dkin)+(\dkjn\!+\!\djkn)} \nnn\hsbb
     \sa= 3 + w\sum_{j=1}^3\sum_{k=1}^3 (d_{jk}+d_{kj}) + w \sum_{k=1}^3 (d_{1k}-d_{k1}). 
\eeqa
In the last step, we use $d_{ii}=0$. In general, abundance~$\xi$ of
strategy~$S_i$ can be calculated similarly;
\beqa \hsbb
 x_i \ap= 3 + w \sum_{j=1}^3\sum_{k=1}^3 \pa{d_{jk}+d_{kj}} + w \sum_{k=1}^3 \pa{d_{ik}-d_{ki}} \nnn\hsbb
     \a=  3 - 2w \theta_s \sn \sum_{j=1}^3\sum_{k=1^3} \pa{\ajj-\ajk-\akj+\akk} \nnn\hsbb
     & & \hsa   +\, 2w \theta_a \sum_{k=1}^3\, \pa{\sn\aii+\aik-\aki-\sn \akk}. 
\eeqa 
Since the first two terms are independent of~$i$, abundance order is
determined by the third term. In other words,
strategy~$S_i$ is more abundant than strategy~$S_j$ when 
\beqa
 \sum_{k=1}^3 \lik  \ag \sum_{k=1}^3 \ljk,
\label{e.cond.w}
\eeqa
where
\beqa
 \lik \a= \sn \aii + \aik - \aki - \sn \akk.
 \label{e.lik2}
\eeqa
Here, inequality~(\ref{e.cond.w}) is derived for abundance with three strategies.
Its generalization with $n$ strategies, 
$\sum_{k=1}^n \lik > \sum_{k=1}^n \ljk$, 
can be derived similarly.  

\subsection{Optional prisoner's dilemma game}
The analysis used in Section~\ref{ss.se} can be
also applied to strategy selection on 
optional prisoner's dilemma game 
[with payoff given by Eq.~(\ref{e.CDLn})].
Let $\Delta x^u$ be the difference between (unnormalized)
abundance of~$C$ and~$D$, i.e., $\Delta x^u = \xc^u - \xd^u$,
where $\xc^u$ and $\xd^u$ are given by Eq.~(\ref{e.eigenCDL}).
Then, cooperators are more abundant than defectors when 
$\Delta x^u$ is positive. 
When $Nw$ is much less than 1, we have 
\beqa \hsbb
\Delta x^u 
  \a=  \pa{\rcl+\rdl} \pa{\rcd-\rdc} + n \pa{\rcl\rld-\rlc\rdl}\nnn \hsbb
  \ap= 2 w \pa{\dcd-\ddc} + n w \pa{\dcl+\dld - \dlc-\ddl}\nnn \hsbb
  \a= 4 w \theta_a \pa{\sn R + S - T - \sn P} 
      + 2nw\theta_a \pas{\sn(R-g)+\sn(g-P)}\nnn\hsbb
  \a= 4 w \theta_a \pas{\frac{2+n}{2}\sn R\, +\, S\, -\, T\, -\, \frac{2+n}{2}\sn R}.
\eeqa
Here $\theta_a$ and $\sn$ are given by Eq.~(\ref{e.sntheta})
and $\dij$ is given by Eq.~(\ref{e.dij}) with payoff
matrix element given by Eq.~(\ref{e.pmCDL}).
Since $\xc>\xd$ when $\Delta x^u$ is positive, we have more
cooperators than defectors when 
\beqa
 \sn(n)\, R + S &>&  T + \sn(n)\, P
\label{e.rstpw2}
\eeqa
with 
\beqa \hsbb
 \sn(n) \sa= \left\{\begin{array}{lll}
  \pa{1 + \frac{1}{2}\,n}\pa{1-\frac{2}{N}}  &\hsbb \mf \mbox{BD\ \&  MP}	\\ 
  \pa{1 + \frac{1}{2}\,n}\pa{3-\frac{8}{N}}  &\hsbb \mf \mbox{DB}.
 \end{array}
 \right.
\eeqa
For large population limit ($N\rightarrow\infty$), $\sn(n)$ becomes
\beqa \hsg
 \sigma(n) \sa= \left\{\begin{array}{lll}
   1+\frac{1}{2}\,n  & \mf \mbox{BD\ \& MP}	\\ 
   3+\frac{3}{2}\,n  & \mf \mbox{DB}.
 \end{array}
 \right.
\eeqa
The structure factor, $\sigma(n)$ becomes $\sigma$ of Eq.~(\ref{e.sn})
when $n=0$ (without loner strategy). Then, 
cooperators are more abundant
than defectors when $R+S>T+P$ for BD \& MP and 
$3R+S > T + 3P$ for DB as expected. 
On the other hand, 
the social dilemma is completely resolved ($\xc>\xd$ whenever $R>P$)
when the number of loner types, $n$, goes to infinity.   

We observe that condition \eqref{e.rstpw2} for the success of
cooperation does not depend on the loner payoff $g$.  This may be
counter-intuitive, since the abundance of loners increases with $g$,
and cooperators fare better when loners increase.  However, in the
$wN$ limit, the frequency of loners is a first-order deviation from
$n/(n+2)$.  The effect of this deviation on cooperators is a
second-order effect that disappears in the $wN$ limit. 

\section{Analysis of the $Nw$ limit}

Here, we consider the results of Section 3 under the $Nw$ limit.
We first calculate fixation probability in the large~$N$ limit 
using Eq.~(\ref{e.fixBD}). The $Nw$ limit is obtained by taking
the $w \to 0$ limit of the result. Once we obtain fixation probability
in this limit, we calculate abundance and find the condition for
the strategy~$S_i$ is more abundant than the strategy~$S_j$ for 
three strategy games. 

\subsection{Fixation probability}
Fixation probability of Eq.~(\ref{e.fixBD}) is is valid for general~$w$
and~$N$ for BD and DB. When $N$ goes to infinity  (with a finite~$w$), 
$\rab$ becomes zero if $y>1$ since 
the $N$th power term in Eq.~(\ref{e.fixBD}) 
becomes infinity. When $y<1$, the $N$th power term becomes zero
and $\rab$ of Eq.~(\ref{e.fixBD}) becomes $f/g$. 
Since $y<1$ when $a+b<c+d$ for BD (and when $3a+b<c+3d$ for DB), 
the fixation probabilities in the limit of large population limit are
given by 
\beqa
 \rab \sa= \frac{e^{w(a+b)}-e^{w(c+d)}}{e^{w(a+b)}-e^{w(c+d)}+e^{w(a-b+c+d)}}
\label{e.NfixBD}
\eeqa
when $a+b > c+d$ and 0 otherwise for BD, 
and 
\beqa \hsg
 \rab\hsg \sa= \hsg \left[ \frac{3+e^{2w(d-b)}}{2} 
    + \frac{e^{w(c+d)}\pa{e^{2wb}+e^{2wd}}\pa{e^{w(a+b)}+e^{2wd}}\pa{e^{2wa}+e^{w(c+d)}}}
           {2e^{2wb}\pa{e^{w(a+b)}+e^{w(c+d)}}\pa{e^{w(3a+b)}-e^{w(c+3d)}}}
           \right]^{-1}
 \nnn\hsbb\hsg
\label{e.NfixDB}
\eeqa
when $3a+b > c+3d$ and 0 otherwise for DB. 
For MP, $\rab$ can be approximated by Eq.~(\ref{e.fixMPi}) for
large~$N$. 

Fixation probability in the $Nw$ limit is obtained by taking 
$w\rightarrow 0$ limit to Eqs.~(\ref{e.NfixBD}) and~(\ref{e.NfixDB}).
In this limit, $\rab$ becomes
\beqa \hsbb\hsg
 \rab \hsg\sa=\hsg  \left\{\begin{array}{ll}\hsg
	w\pa{\,a+b-c-\,d}\Theta(\,a+b-c-\,d) &\hsbb \mf  \mbox{BD}	\\ \hsg
	w\pa{3a+b-c-3d}  \Theta(3a+b-c-3d)   &\hsbb \mf  \mbox{DB}.	 
	\end{array}
	\right.  
\label{e.fix.liss}
\eeqa

This result can be also understood from random walk argument on 1D
lattice. Here, $Nw$ is much larger than 1 and hence diffusion to 
drift-velocity ratio, $d/v \approx 1/Nw$ is small.  Hence, population dynamics is
mainly determined by the (biased) drift term rather than the
stochastic diffusion. Fixation (random walker at $\na=N$ state) is now possible 
only when the drift bias is positive for
(almost) everywhere.  For BD and DB on cycles, drift velocity is
independent of~$\na$ and proportional to $\sigma a + b - c - \sigma d$. 

\subsection{Strategy selection} 
We now consider the condition for $\xi>\xj$ 
in the large population limit with finite $w$ for BD and DB. 
As mentioned before, we are comparing abundance $\xi$ and $\xj$ 
in the population with three strategies, $S_i$, $S_j$ and $S_k$. 
We first note that $\xj$ and $\xk$ are zero 
when both $\rji$ and $\rki$ are zero [see Eq.~(\ref{e.eigen})].
This is the case when both  $\lji$ and $\lki$ are negative [see
Eq.~(\ref{e.rikNw})] where $\lij = \sigma \aii + \aij - \aji - \sigma \ajj$.
Therefore, $1 = x_i > x_j = 0$ if both $\lij$ and $\lik$ are positive. 
By the same token, $0=x_i < x_j = 1$ when both $\lji$ and $\ljk$ are positive.
If $\lki$ and $\lkj$ are positive, both $\xi$ and $\xj$ are zero.
Hence, the condition for $\xi>\xj$ becomes non-trivial
only when three strategies show rock-paper-scissors characteristics.
For the $\lij>0$ case (with $\ljk>0$ and $\lki>0$),
$\xi^u$ and $\xj^u$ in Eq.~(\ref{e.eigen}) become
$\rij\rjk$ and $\rjk\rki$ respectively.
Therefore, $S_i$ is more abundant than $S_j$ when $\rij>\rki$.
For the other case of $\lji>0$ (with $\lik>0$ and $\lkj>0$),
$\xi^u$ and $\xj^u$ become $\rik\rkj$ and $\rji\rik$ 
and $\xi > \xj$ when $\rkj>\rji$.
Hence, there are three cases that strategy~$S_i$ is more abundant than
strategy~$S_j$ in the large population limit; 
\bit
 \item case 1\, [$\lij>0$ and $\lik>0$]:\ $\xi>\xj$ always,
 \item case 2\, [$\lij>0$, $\ljk>0$, and $\lki>0$]:\ $\xi>\xj$ if  $\rij>\rki$, and
 \item case 3\, [$\lji>0$, $\lik>0$, and $\lkj>0$]:\ $\xi>\xj$ if $\rji<\rkj$.
\eit
For the cases~2 and~3, 
conditions for $\xi>\xj$ can be understood by integrating out the
role of strategy~$S_k$.  
For the case~2, influx to strategy~$S_i$ is $\rij \xj$  while out-flux is $\rki \xi$. 
Therefore, detailed balance between the abundance of~$S_i$ and~$S_j$ in the steady state 
requires
\beqa
 \rij \xj \a= \rki \xi.
 \label{e.case2}
\eeqa
Hence, $\xi = \frac{\rij}{\rkj}\, \xj$ is
larger than $\xj$  if $\rij>\rki$.
For the case~3, influx to strategy~$S_i$ 
is $\rkj \xj$ when the role of strategy~$S_k$ is integrated out.
Since the out-flux to strategy~$S_i$ is $\rji \xi$, we have
\beqa
 \rkj \xj \a= \rji \xi
 \label{e.case3}
\eeqa
in the steady state, and $\xi = \frac{\rkj}{\rji}\, \xj$ is
larger than $\xj$  if $\rkj>\rji$.
From the large $N$ limit of $\rij$ in Eq.~(\ref{e.fixBD}), 
we see that the conditions for $\xi>\xj$ for the
cases~2 and~3 become
\beqa
\begin{array}{llll}
 \fij\,\gki &>& \fki\,\gij  &\mf \mbox{case~2} \\ 
 \fkj\,\gji &>& \fji\,\gkj  &\mf \mbox{case~3}.
\end{array}
\eeqa
Here
\beqa
 \fij \a= \alii\alij-\alji\aljj\nnn
 \gij \a= \alii\alij-\alji\aljj + \alii\alij^{-1}\alji\aljj
\eeqa
for BD, and 
\beqa \hsbb\hsg
 \fij \sa= \alii^{3}\alij-\alji\aljj^{3}\nnn\hsbb\hsg
 \gij \sa=  \frac{3\fij+\alij^{-1}\aljj^2\fij}{2}\!
        +\! \frac{\alji\aljj\pa{\alij^2+\aljj^2}
          \pa{\alii\alij+\aljj^2}\pa{\alii^2+\alji\aljj}}
              {2\alij^2\pa{\alii\alij+\alji\aljj}}
\label{e.fijgij}
\eeqa
for DB with $\alij = e^{w\aij}$. 

Now we consider the $Nw$ limit, where $w$ goes to zero after $N$ goes
to infinity.
In this case, $\fij$ and $\gij$ in Eq.~(\ref{e.fijgij}) become 
linear in $w$ and $\rij$ becomes proportional to
$\lij$ (unless $\lij<0$ where $\rij=0$).
The conditions for three cases for large population become
\bit
 \item case 1\, [$\lij>0$ and $\lik>0$]:\ $\xi>\xj$ always.
 \item case 2\, [$\lij>0$, $\ljk>0$, and $\lki>0$]:\ $\xi>\xj$ if $\lij>\lki$.
 \item case 3\, [$\lji>0$, $\lik>0$, and $\lkj>0$]:\ $\xi>\xj$ if $\lji<\lkj$.
\eit

\subsection{Optional prisoner's dilemma game}
We now consider optional prisoner's dilemma game whose payoff matrix
is given by Eq.~(\ref{e.CDLn}). We first assume $R>P$. 
In general, the effect of loners on the
strategy selection between~C and~D disappears if $R=P$ due to the symmetry. 
Hence, we need to consider $R\not=P$ case only and assume 
$R>P$ without loss of generality.
We further assume that $R>g$. Otherwise, 
both $\lcl = \sigma (R-g)$ and $\ldl = \sigma (P-g)$
are negative and both $\xc$ and $\xd$ become 0.
When we assume $R>P$ and $R>g$, 
two possibilities are left, 
$P>g$ and $P<g$.

As before, we consider the difference
between $\xc^u$ and $\xd^u$ [given by Eq.~(\ref{e.eigenCDL})]
and let $\Delta x^u = \xc^u - \xd^u$. 
When $g<P$, both $\rlc$ and $\rld$ are zero
since both $\llc$ and $\lld$ are negative
and we get 
\beqa
 \Delta x^u \a= \pa{\rcl+\rdl} \pa{\rcd-\rdc}
\eeqa 
from Eq.~(\ref{e.eigenCDL}). Therefore, $\xc>\xd$ 
when 
\beqa
 \rcd \ag \rdc.
\label{e.ssxca}
\eeqa
This can be easily understood since abundance of loners becomes zero 
when $Nw\gg 1$ in the $g<P$ case.
On the other hands, for the $g>P$ case, 
$\Delta x^u$ becomes
\beqa
 \Delta x^u \a=\rcl\pa{\rcd + n \rld - \rdc}.
\label{e.dxuss}
\eeqa
Therefore, $\xc > \xd$ when 
\beqa
 \rcd  \ag \rdc - n\rld.
\label{e.ssxcb}
\eeqa

The inequalities~(\ref{e.ssxca}) and~(\ref{e.ssxcb}) 
are valid as long as $Nw$ is much larger than 1 for general~$w$.
There are three possibilities for $Nw$ to go infinity, $w$ goes to 
infinity, $N$ goes to infinity or both go to infinity. 
Let us first consider the $Nw$ limit in which $N \to \infty$ first and
then $w \to 0$.
In this case, the conditions for $\xc>\xd$ on cycles,
inequalities~(\ref{e.ssxca}) and~(\ref{e.ssxcb}) can be written as
linear inequalities.   Here, $\rcd-\rdc$ is always proportional to
$\lcd$.  Also, $\rld$ becomes proportional to $\lld$ if $g>P$.  
Therefore, we have $\xc>\xd$ when
\beqa \hsbb
 \sigma R + S  &\hsg >\hsg & T + \sigma P - n\sigma (g-P)\Theta(g-P), 
\label{e.rstps}
\eeqa
where $\sigma=1$ for BD and 3 for DB. 

Now, let us consider high intensity of selection limit
where $w$ itself goes to infinity. Then, $\rld$ becomes 1 when $g>P$
since loners dominates defectors and inequality~(\ref{e.ssxcb}) becomes
\beqa
 \rcd  \ag \rdc - n.
\eeqa
This implies that cooperators are more abundant than defectors
always for large $w$ if $n>1$ since $\rdc$ cannot be larger than 1. 

\section{Optional game with simplified prisoner's dilemma}

To further clarify how spatial structure and optionality of the game
affect the success of cooperation, we study a optional version of a
simplified prisonser's dilemma, in which cooperators pay a cost $c$ to
generate a benefit $b$ for the other player.  This simplified
prisoner's dilemma is also known as the donation game or the
prisoner's dilemma with equal gains from switching.  
Here, we consider the $n=1$ optional game with a simplified 
prisoner's dilemma, whose payoff matrix is given by 
\beqa
\begin{array}{cl}
     & \begin{array}{ccc}
       \hsb C  & \hsc\ \ D & \hsb L 
       \end{array} \ \	\\
\begin{array}{c}
     C\hsg \\
     D\hsg \\
     L\hsg 
\end{array}
   & \left( \begin{array}{ccc}
   b-c   &\hsb -c & \hsb\ g \\
   b   &\hsb 0 & \hsb\ g \\ 
   g   &\hsb g & \hsb\ g  
  \end{array} \right).  
\end{array}
\label{e.pmSPD}
\eeqa
Here, $g$ is the payoff for a loner (for staying away from a game) and
$b$ and $c$ are the benefit and cost of the cooperation respectively.
We assume that the cost to participate the game, $g$, is positive but
less than the benefit of cooperation and consider parameter
regions of $0<c<b$ and $0<g<b$. 

For the simplified PD game, we have $R=b-c$, $S=-c$, $T=b$ and $P=0$ 
and the condition for $\xc>\xd$ in the $wN$ limit, 
given by inequality~(\ref{e.rstpw2}), becomes
\beqa
  c &<&\frac{N-6}{5N-6} \, b
    \b= \frac{b}{5} + O(1/N)
\label{ie.ws.bd}
\eeqa
for BD and MP, and 
\beqa
  c &<&\frac{7N-24}{11N-24} \, b
   \b=  \frac{7b}{11} + O(1/N)
\label{ie.ws.db}
\eeqa
for DB. Note that the condition for $\xc>\xd$ is independent of $g$, as we saw earlier in Section 4.3.
In the $wN$ limit, the condition for $\xc>\xd$ mainly depends on the frequency
of loners, which is roughly 1/3 regardless of $g$ values.

On the other hand, in the $Nw$ limit, inequality~(\ref{e.rstps}) becomes 
\beqa
 g & > & 2c
\label{ie.ss.bd}
\eeqa
for BD and
\beqa
 g & > & \frac{4}{3}\pa{c-b/2} 
\label{ie.ss.db}
\eeqa
for DB.

Now we consider the large $w$ limit ($w\gg 1$).
First, note that the condition for $\xc > \xd$, 
given by inequality~(\ref{e.ssxcb}), becomes
\beqa
 \rcd  \ag \rdc - \rld
\label{e.ssxcc}
\eeqa
when $n=1$. The fixation probabilities, $\rcd$, $\rdc$ and $\rld$ can be
easily calculated from Eq.~(\ref{e.fixBD}) for large $w$. 
For BD,  $\rcd$, $\rdc$ and $\rld$ become
$e^{-cNw}$, $1-e^{-(b+2c)w}$ and $1-e^{-gw}$ respectively 
for sufficiently large~$w$ and inequality~(\ref{e.ssxcc}) becomes
\beqa
 e^{-cNw}  \ag e^{-gw} - e^{-(b+2c)w}.
\eeqa
Since, $e^{-gw}$ is larger than $e^{-(b+2c)w}$ when $g<b$,
inequality~(\ref{e.ssxcc}) cannot be satisfied for large population ($N>g/c$).
In other words, $\xd$ is always larger than $\xc$ for BD in the $w\rightarrow\infty$ limit.
It is worthwhile to note how strongly 
strategy selection depends on the number of loner types for large~$w$. 
As discussed before, 
cooperators are more abundant than defectors if 
the types of loners, $n$ is larger than 1. 
On the other hand, for $n=1$, defectors are more abundant than cooperators
as long as $0<g<b$.

For DB, we get similar results for $\rcd$ and $\rld$. 
As $w$ goes to infinity, $\rcd$ becomes zero while $\rld$ becomes $2/3$. 
On the other hand, $\rdc$ depends on the 
benefit to cost ratio. It is $2/3$ if $c$ is larger than $b/2$ and
zero otherwise. 
Hence, cooperators are more abundant than defectors when $c<b/2$.

For MP, we calculate fixation probabilities directly 
using Eq.~(\ref{e.fix}) in the limit of $w\rightarrow\infty$ 
and find that  $\rcd/\rdc$ becomes 
$1 + e^{-wc} - e^{-wg}$ for large~$w$. 
Hence, cooperators are more abundant than defectors when $g>c$.  

This simplified game allows us to examine how spatial structure and optionality of the game combine to support cooperation.

\section{Numerical analysis}
 \label{s.na}
We have analyzed the conditions for strategy selection analytically
in the two extreme limits of selection intensity, $w\to0$ and $w\to\infty$
in the zero mutation rate.
Here, we first we obtain conditions for $\xc>\xd$ in the simplified game \eqref{e.pmSPD} numerically 
for finite values of~$w$ (with low mutation rate),
using calculated abundance from fixation probabilities. 
Then, we perform a series of Monte Carlo simulations 
with small but finite mutation rates. 
The condition for strategy selection is obtained numerically 
using measured abundance in the simulations. 

\subsection{Numerical comparison of abundance of cooperators and defectors}
\begin{figure}[!t]
\includegraphics[width=.5\textwidth,keepaspectratio=true]{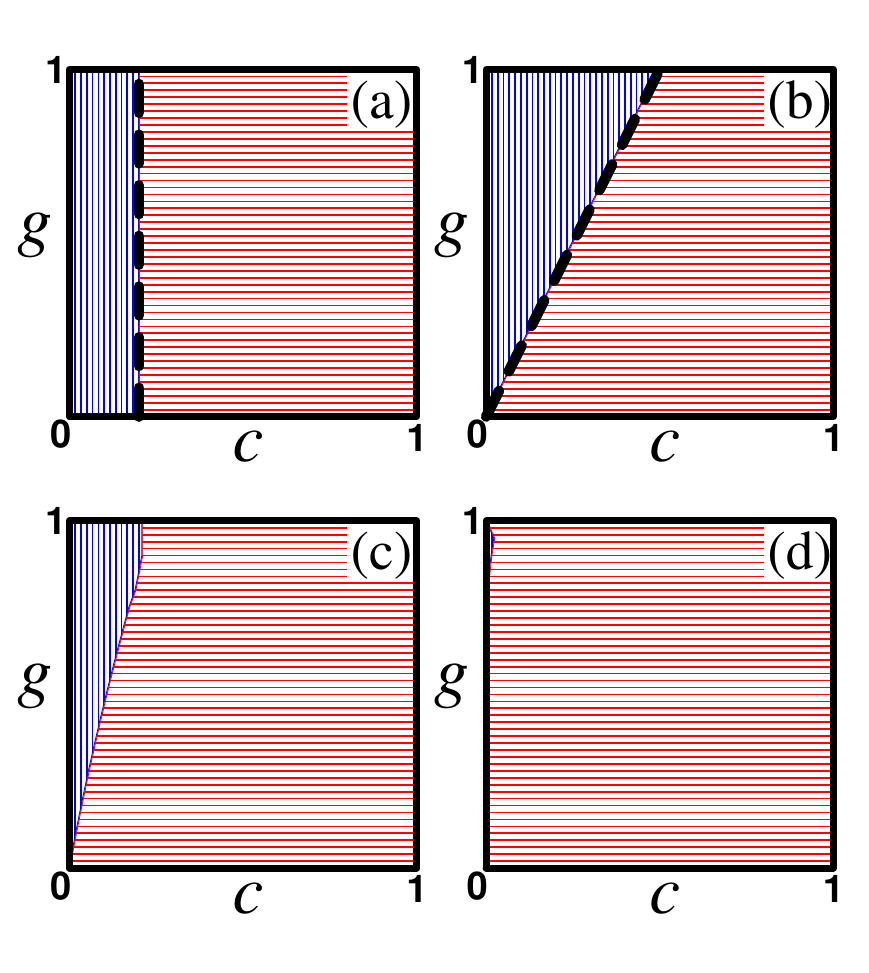}
\caption{
C-rich (blue-vertical) and D-rich (red-horizontal) regions 
for BD in the $c$-$g$ parameter space. 
Population size is $N=10^4$ and selection intensities are 
(a) $w=10^{-6}$, (b) $w=10^{-2}$, (c) $w=1$, and (d) $w=10$.
Black lines in~(a), and~(b) are given by $c=1/5$ and $g=2c$
respectively. 
} 
\label{f.phaseBD}
\end{figure}
\begin{figure}[!t]
\includegraphics[width=.5\textwidth,keepaspectratio=true]{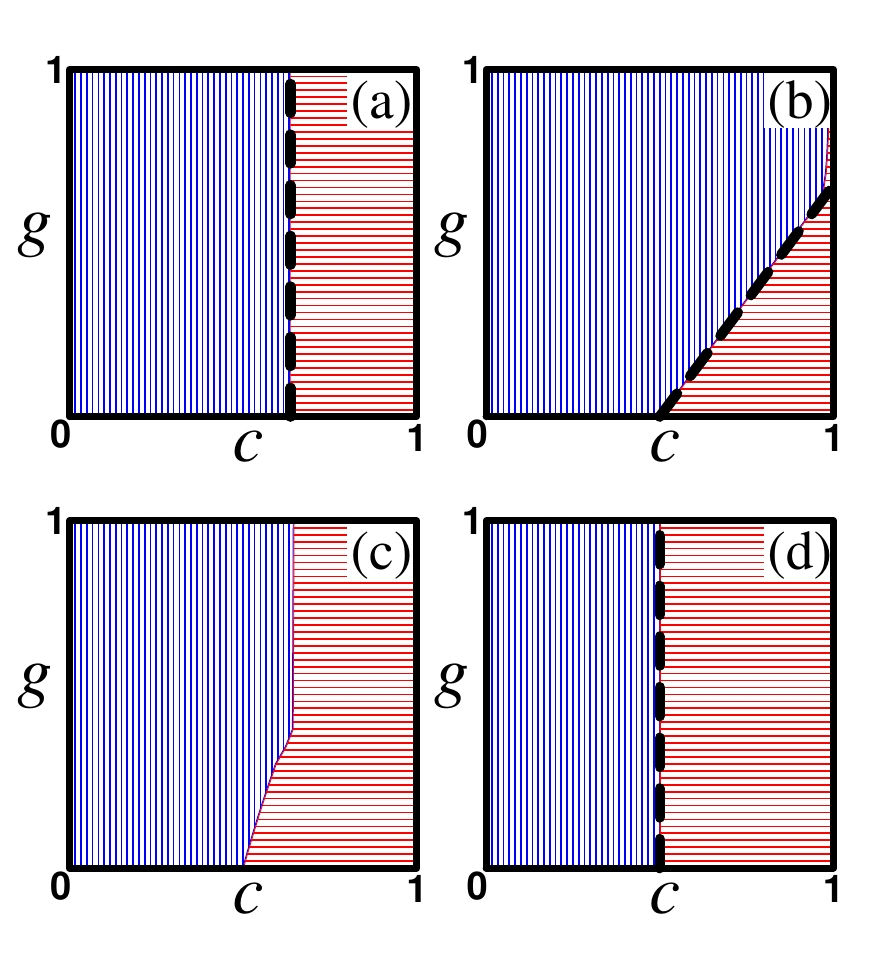}
\caption{
C-rich (blue-vertical) and D-rich (red-horizontal) regions 
for DB in the $c$-$g$ parameter space. 
Population size is $N=10^4$ and selection intensities are 
(a) $w=10^{-6}$, (b) $w=10^{-2}$, (c) $w=1$, and (d) $w=10$.
Black lines in~(a), (b), and~(d) are given by $c=7/11$ 
$g=\frac{4}{3}\pa{c-\half}$, 
and $c=1/2$ respectively. 
} 
\label{f.phaseDB}
\end{figure}
\begin{figure}[!t]
\includegraphics[width=.5\textwidth,keepaspectratio=true]{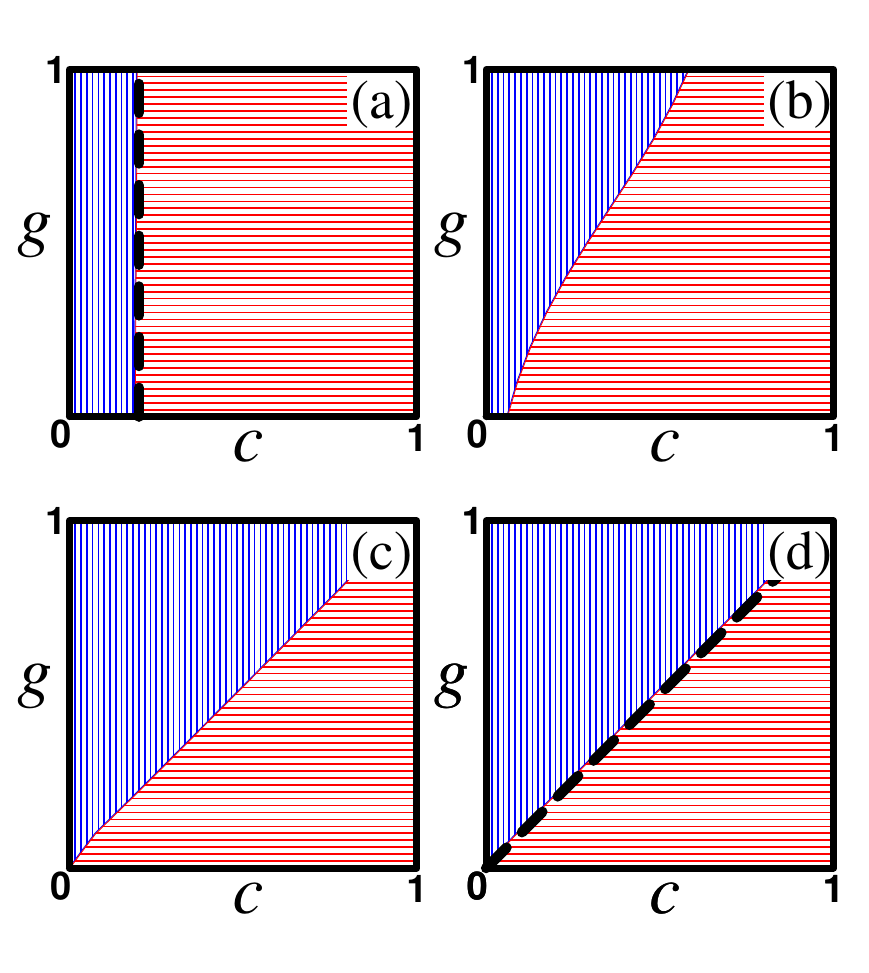}
\caption{
C-rich (blue-vertical) and D-rich (red-horizontal)
regions for MP in the $c$-$g$ parameter space. 
Population size is $N=100$ and selection intensities are 
(a) $w=10^{-3}$, (b) $w=10^{-1}$, (c) $w=1$, and (d) $w=10$.
Black lines in~(a) and~(d) are given by $c=1/5$ 
and $g=c$ respectively. 
} 
\label{f.phaseMP}
\end{figure}

We solve the inequality $\xc > \xd$ numerically 
using abundance given by Eq.~(\ref{e.eigenCDL})
with $n=1$ and investigate how the boundaries between C-rich and
D-rich regions in the parameter space change as the selection intensity,~$w$ varies. 
Without loss of generality, we set $b=1$ and investigate the parameter
space given by  $0<c<1$ and $0<g<1$. The boundaries are obtained by
finding $c$ which satisfies $\xc=\xd$ for a given $g$.  

In Fig.~\ref{f.phaseBD}, we draw C-rich and D-rich regions 
for BD by blue-vertical and red-horizontal lines respectively 
for four different values of selection intensities.
C-rich regions in~(a) and~(b) are consistent with 
the analysis in the 
$wN$ limit [inequality~(\ref{ie.ws.bd})]
and in the $Nw$ limit [inequality~(\ref{ie.ss.bd})] respectively.
The dark-dashed lines, given by $c=1/5$ and $g=2c$, are the boundaries
between C-rich  and D-rich regions predicted in the $wN$ and $Nw$
limits respectively.
For $w=10$ shown in~(d), defectors are more abundant for
almost entire region. This is consistent with 
the $w\rightarrow\infty$ analysis
which always predict $\xd>\xc$ for $n=1$. 
For the intermediate value of $w=1$ shown in~(c), we
do not know the analytic boundary but 
we observe that the numerical boundary lies between  
the boundary for $w=10^{-2}$ of~(b) and
that for $w=10$ of~(d) as expected. 

For DB, we show C-rich and D-rich regions for $N=10^{4}$  
in Fig.~\ref{f.phaseDB}. As in Fig.~\ref{f.phaseBD}, they are 
represented by blue-vertical and red-horizontal lines respectively for
four different values of selection intensities.
C-rich regions in~(a) and~(b) coincide with the predictions
for the $wN$ and $Nw$ limits respectively. 
The dark-dashed lines, given by $c=7/11$ and
$g=\frac{4}{3}\pa{c-\half}$, are the boundaries 
between C-rich and D-rich regions predicted in the $wN$
and $Nw$ limits respectively.
For $w=10$ shown in~(d), cooperators are more abundant 
if $c<1/2$ as predicted in the $w\rightarrow\infty$ limit.
As in the case of BD, we do not know the analytic
boundary for the intermediate value of $w=1$ shown in~(c). 
Yet, at least, we confirm that the numerical boundary lies between  
the boundary in the $Nw$ limit and that in the $w\rightarrow\infty$ limit.

In Fig.~\ref{f.phaseMP}, we show C-rich and D-rich regions 
for MP by blue-vertical and red-horizontal lines
respectively. For MP, we do not have an analytic expression for the 
fixation probability in a closed form. Hence we need to 
calculate fixation probabilities directly from 
Eq.~(\ref{e.fix}). 
Due to numerical cost for calculating abundance, which
increases rapidly with $N$, we investigate relatively small 
population of $N=100$. However, they seem to be big enough to confirm
the analytic prediction of the boundaries between C-rich and D-rich
regions in the $wN$ limit and in the large $w$ limit. 
The dark-dashed lines in~(a) and~(d), given by $c=1/5$
and $g=c$, are the predicted boundaries in 
the $wN$ and large $w$ limits respectively. 

\subsection{Combined effects of optionality and spatial structure}

Now, let us compare the effects of the option to  be loners  on the
structured  population  (BD  and  DB)   to  those  on  the  well-mixed
population (MP).  It is immediately clear that the effects of spatial
structure depend on the update rule. 
Comparing Figures 1 and 3, we see that BD updating does not support
cooperation, in accordance with findings from other models 
\citep{Ohtsuki:2006gp,Ohtsuki:2006ug,Debarre:2014tq}
In panels 1(a) and 3(a),
where $Nw=0.1$, the C-rich regions for BD and MP appear to coincide.
This accords with our results that, in the $wN$ limit, the condition
for $x_C>x_D$ is $c<b/5$ for both MP and BD (see Section~6).  In the
other panels of Figures 1 and 3, we see that the C-rich regions for BD
are smaller than those for MP, suggesting that BD updating actually
impedes cooperation relative to its success in a well-mixed
population.  

DB updating is generally favorable to cooperation, as can be seen by
comparing Figures 2 and 3.  In the $wN$ limit, for example, the
condition for $x_C>x_D$ is $c<7b/11$ under DB updating (see Section
6), which is less stringent than the corresponding condition for MP,
$c<b/5$.  These conditions correspond approximately to the C-rich
regions shown in Figrues 2(a) and 3(a).  However, we find that as $w$
increases, the C-rich regions for DB do not necessarily contain those
for MP.  In other words, for large selection intensity, there are
parameter combinations under which cooperation is favored in a
well-mixed population but disfavored on the cycle with DB updating.
This effect is most visible in Figures 2(d) and 3(d), but it can also
be seen in 2(c) and 3(c).  In the $w \to \infty$ limit, we found
(Section 6) that cooperation is favored for MP if $c<g$, while it is
favored for DB for $c<b/2$.  Either one of these conditions can be
satisfied while the other fails, as can be seen (approximately) in
Figures 2(d) and 3(d). 

Optionality of the game and spatial structure (with DB updating) are
two mechanisms that support cooperation.  Do these mechanisms combine
in a synergistic way?  We find little evidence that they do.  Let us
consider first the $wN$ limit.  With spatial structure alone (DB
updating with $n=0$ loner strategies), cooperation succeeds if
$c<b/2$.  With optionality alone (MP with $n=1$), cooperation succeeds
if $c<b/5$.  With both optionality and spatial structure (DB  with
$n=1$), the condition is $c<7b/11$, and we observe that the $7b/11$
threshold is less than the sum $b/2 + b/5 = 7b/10$ of the thesholds
corresponding to the the two mechanisms acting alone.  The lack of
synergy is even more apparent as the selection intensity $w$
increases, since, as noted above, there are parameter combinations for
which cooperation is favored for MP but disfavored for DB. 

\subsection{Effects of selection intensity}

\begin{figure}[!t]
\includegraphics[width=.5\textwidth,keepaspectratio=true]{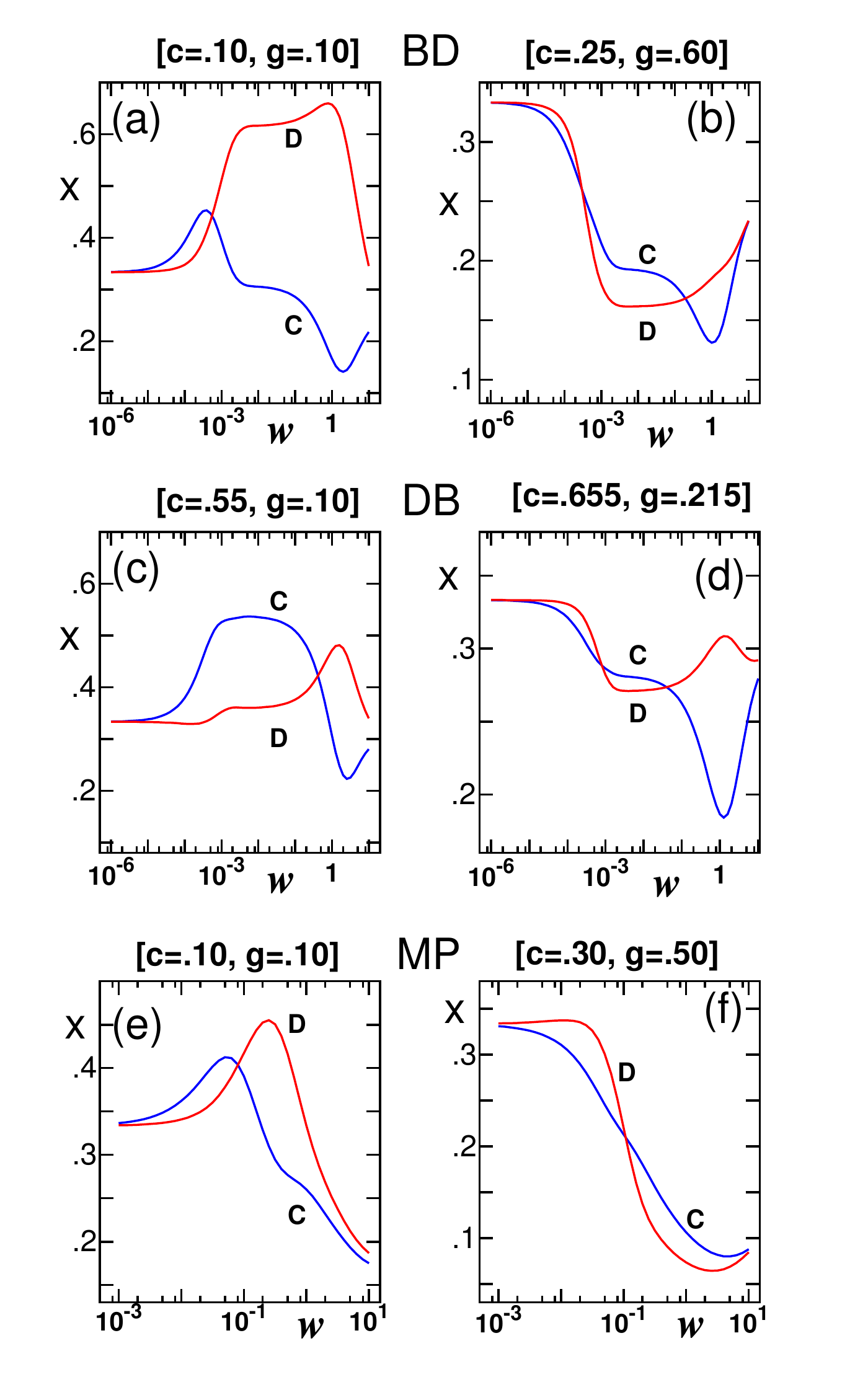}
\caption{
Selection intensity, $w$, dependence of 
abundance, $x$, of cooperators (blue) and defectors (red) 
for BD [(a) and (b)], DB [(c) and (d)], and MP [(e) and (f)]. 
Abundance is numerically calculated using Eq.~(\ref{e.eigen})
with $N=10^4$ for BD and DB, and $N=100$ for MP.
The benefit of cooperation, $b$, is 1.
The costs for a game and a cooperative play, 
denoted by $g$ and $c$ respectively, are shown in the figures.  
Selection intensity~$w$ [$x$-axis] is shown in a log scale 
while abundance~$x$ [$y$-axis] is shown in a linear scale. 
Abundance of loners (not shown) is given by $\xl=1-\xc-\xd$.
}
\label{f.xw}
\end{figure}
\begin{figure}[!h]
\includegraphics[width=.45\textwidth,keepaspectratio=true]{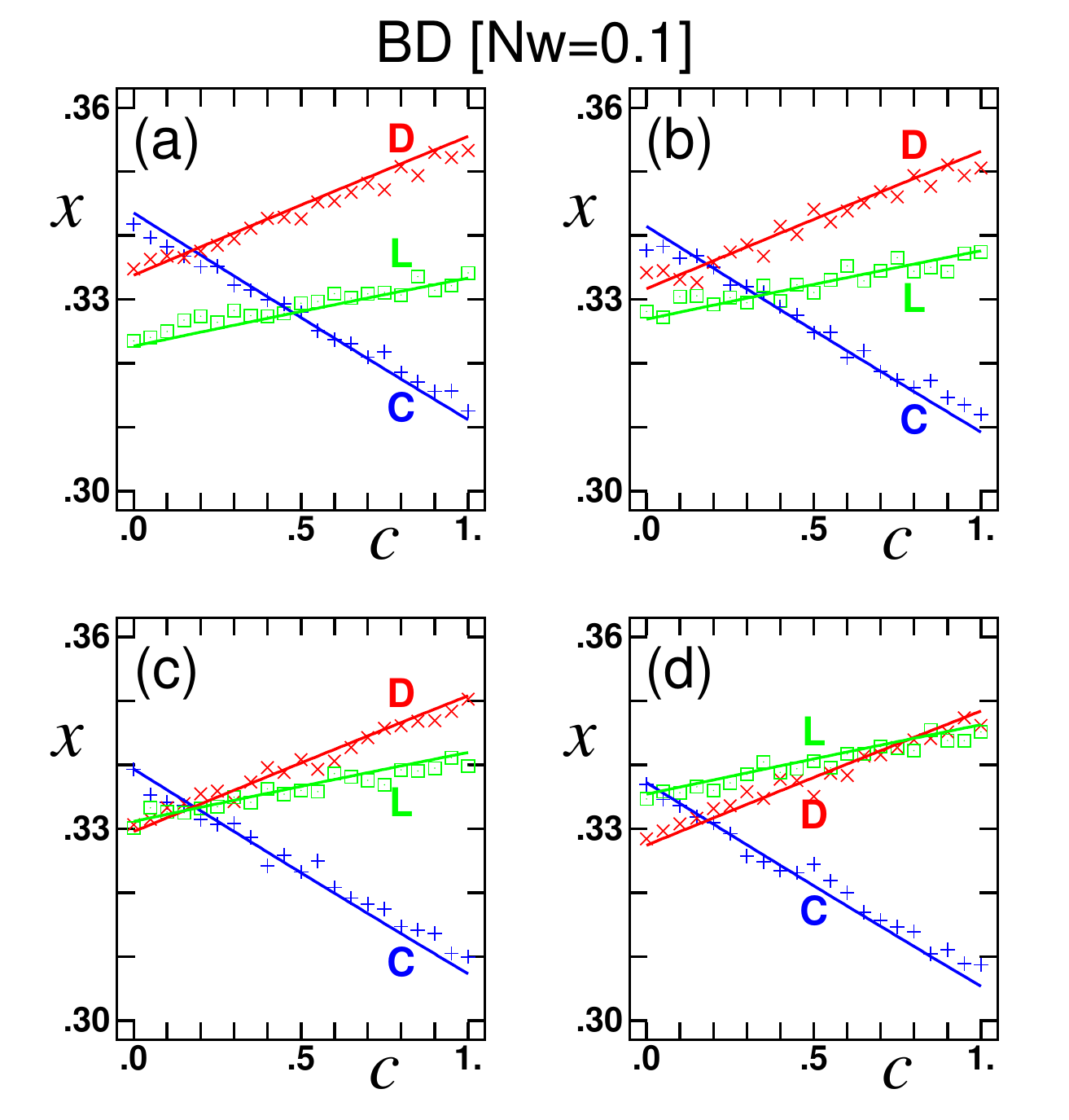}
\caption{
Abundance $\xc$, $\xd$, and $\xl$ vs. $c$ for BD with $N=50$
and $w=0.002$ for four different values of $g$, (a) 0, (b) 0.2, (c)
0.4, and (d) 0.6. Blue plus, red cross, and green square symbols 
represent the $\xc$, $\xd$, and $\xl$ respectively. Blue, red, and
green solid lines are abundance of Eq.~(\ref{e.eigen}).
Mutation rate is $u=0.0002$.
} 
\label{f.xcBD1}
\end{figure}
\begin{figure}[!h]
\includegraphics[width=.45\textwidth,keepaspectratio=true]{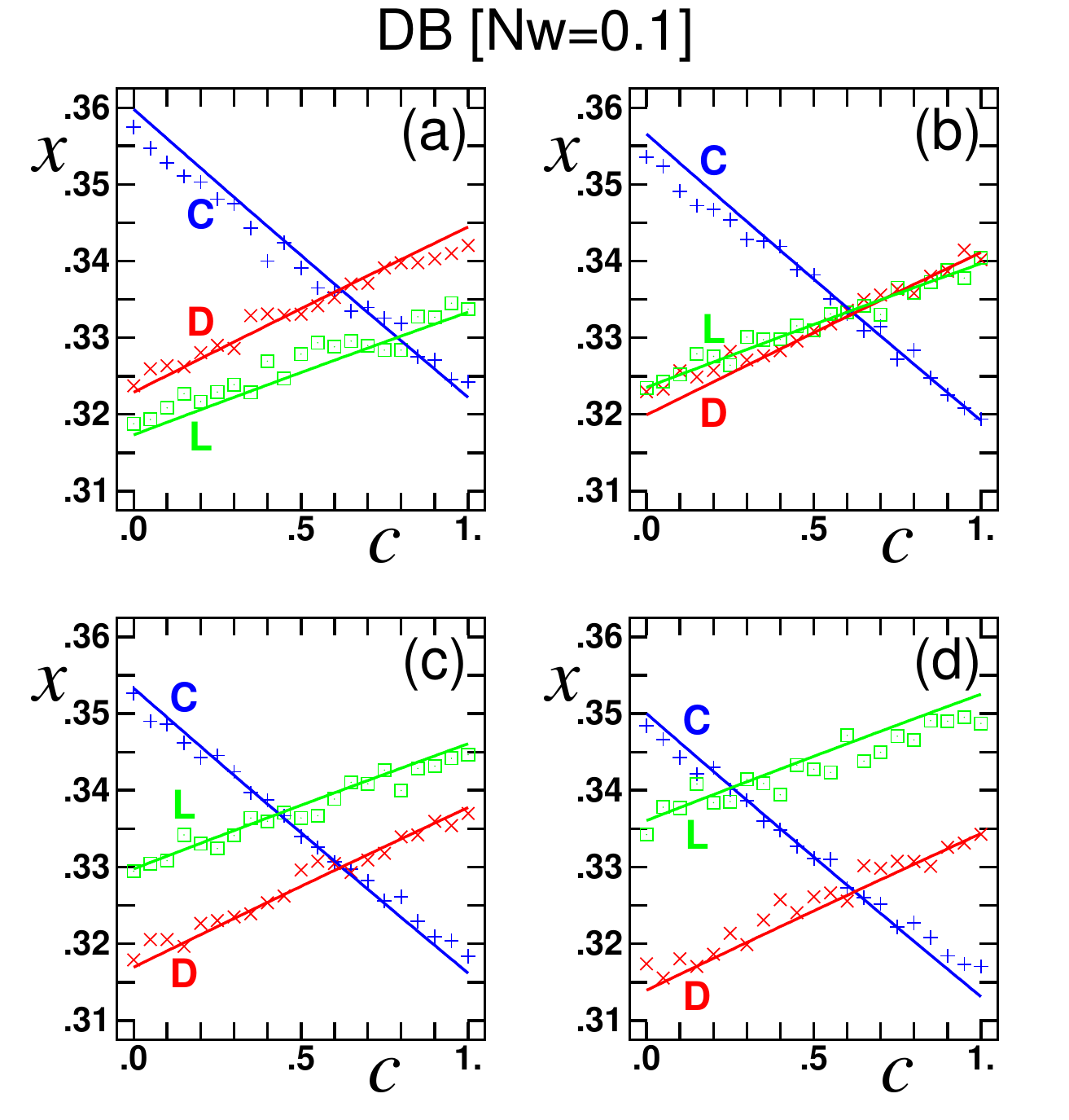}
\caption{
Abundance $\xc$, $\xd$, and $\xl$ vs. $c$ for DB with $N=50$
and $w=0.002$ for four different values of $g$, (a) 0, (b) 0.2, (c)
0.4, and (d) 0.6. Blue plus, red cross, and green square symbols 
represent the $\xc$, $\xd$, and $\xl$ respectively. Blue, red, and
green solid lines are abundance of Eq.~(\ref{e.eigen}).
Mutation rate is $u=0.0002$.
} 
\label{f.xcDB1}
\end{figure}

Let us now take a closer look at the effects of selection intensity.
As shown in Fig.~1-3, the boundary between  $C$-rich region and
$D$-rich region changes  as the selection intensity,~$w$ varies.
In other words, selection intensity may switch 
the rank of strategy abundance for some regions of 
parameter space as recently reported~\citep{Wu:2013uh}.
In Fig.~\ref{f.xw}, we show selection intensity dependence
of abundance for a couple of different pairs of~$c$ and~$g$. Abundance
is numerically  calculated using Eq.~(\ref{e.eigen}) with $N=10^4$ for  
BD and DB. For MP, we consider $N=100$ due to numerical cost. 
In the left panels, we choose parameters~$c$ and~$g$ such that
cooperators are more abundant than defectors ($\xc>\xd$) 
in the $wN$ limit
but change abundance order ($\xd > \xc$) in the $Nw$ limit (for BD and DB)
or large $w$ limit (for MP). 
For (a) BD, (c) DB, and (e) MP, we choose 
$(c,g)=(0.1,0.1)$, $(0.55,0.1)$, and 
$(0.1,0.1)$ respectively
and find ``crossing intensity'',~$w_c$. 
Population remains as $C$-rich phase
for $w<w_c$ where $w_c$ is around 
0.0005, 0.4, and 0.08 for (a), (c), and (e) respectively. 
In the right panels, we consider the opposite cases
and choose parameters such that defectors are more abundant 
in the $wN$ limit
but becomes less abundant in the $Nw$ limit (for BD and DB)
or large $w$ limit (for MP). 
For (b) BD, (d) DB, and (f) MP, we choose 
$(c,g)=(0.25,0.6)$, $(0.655,0.215)$, and 
$(0.3,0.5)$ respectively.
For BD and DB, cooperators seem to be more abundant 
only in the $Nw$ limit. They are less abundant than defectors for
large $w$ limit as well as in the $wN$ limit. In other words, there
are two crossing intensities,~$w_{c_1}$ and~$w_{c_2}$, 
such that $\xc$ is larger than $\xd$ only for $w_{c_1} < w < w_{c_2}$. 
They are given by 
$w_{c_1} = 0.0003$ and $w_{c_2} = 0.25$ for (b)
and $w_{c_1} = 0.001$ and $w_{c_2} = 0.04$ for (d).
For MP shown in~(f), there seems to be only one crossing point 
around at $w=0.1$. 

\begin{figure}[!h]
\includegraphics[width=.45\textwidth,keepaspectratio=true]{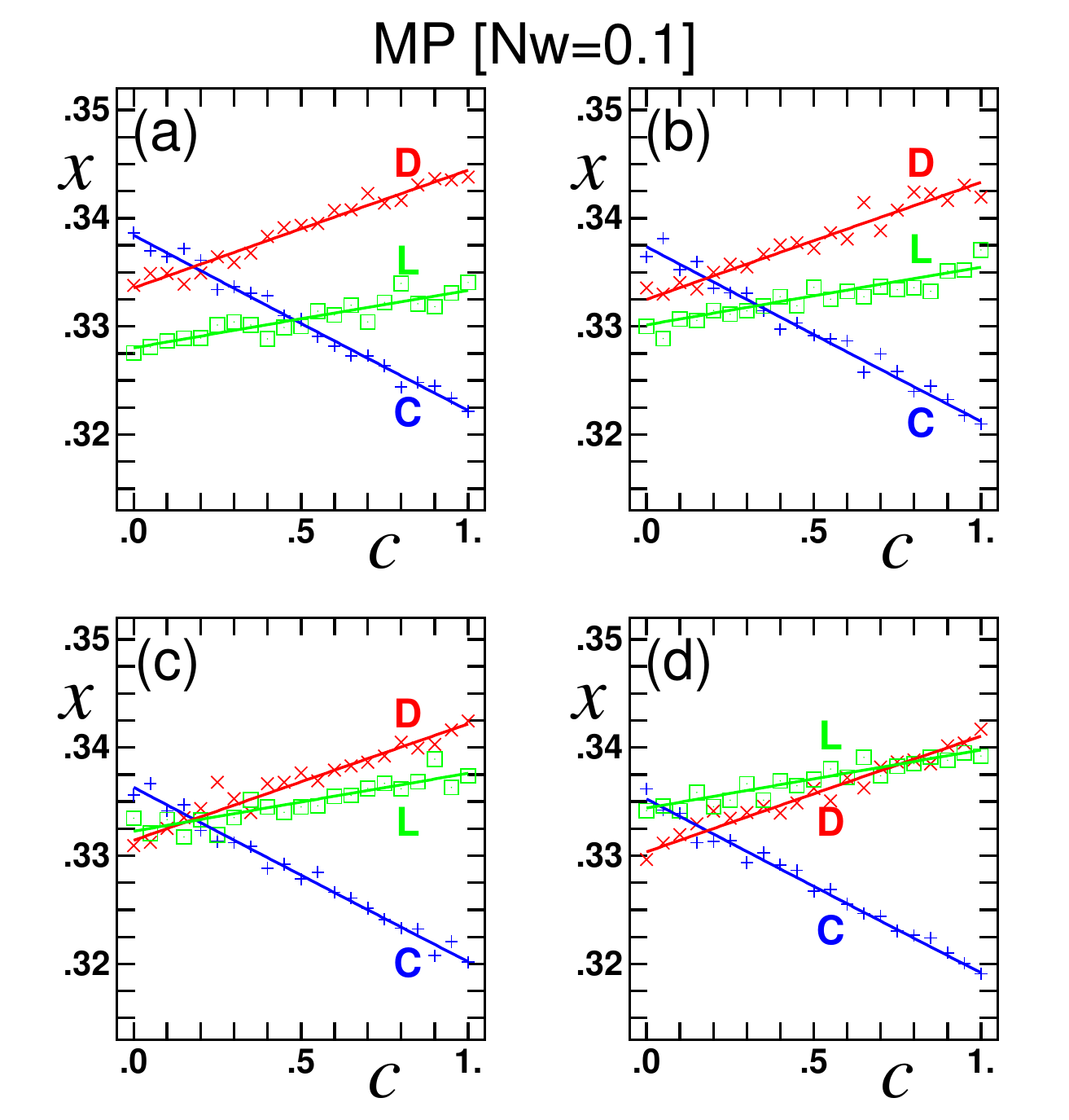}
\caption{
Abundance $\xc$, $\xd$, and $\xl$ vs. $c$ for MP with $N=50$
and $w=0.002$ for four different values of $g$, (a) 0, (b) 0.2, (c)
0.4, and (d) 0.6. Blue plus, red cross, and green square symbols 
represent the $\xc$, $\xd$, and $\xl$ respectively. Blue, red, and
green solid lines are abundance of Eq.~(\ref{e.eigen}).
Mutation rate is $u=0.0002$.
} 
\label{f.xcMP1}
\end{figure}

\subsection{Simulation with finite mutation rate}
Abundance of Eq.~(\ref{e.eigen}) is calculated in the low mutation limit
using the fixation probabilities. 
After the invasion of a mutant to the mono-strategy population, the possibility 
of further mutation during the fixation is ignored.
Strictly speaking, this is valid only 
when the mutation rate $u$ goes to zero. 
Here, we measure the abundance of three strategies,
$\xc$, $\xd$, and $\xl$ by  Monte Carlo simulations 
with a small but finite mutation rate 
and compare them with abundance of Eq.~(\ref{e.eigen}). 

We start from a random arrangement of three strategies
$C$, $D$, and $L$ on a cycle (BD and DB)
or a complete graph (MP) with $N$ sites.  
Population evolves with BD, DB, or MP updating. 
The mutation probability of the offspring is $u$; it
bears its parent strategy with probability $1-u$ 
and takes one of the other two strategies with probability~$u$.
In the mutation process, both strategies have
equal chances, i.e., probability of $u/2$ for each. 

\begin{figure}[!h]
\includegraphics[width=.45\textwidth,keepaspectratio=true]{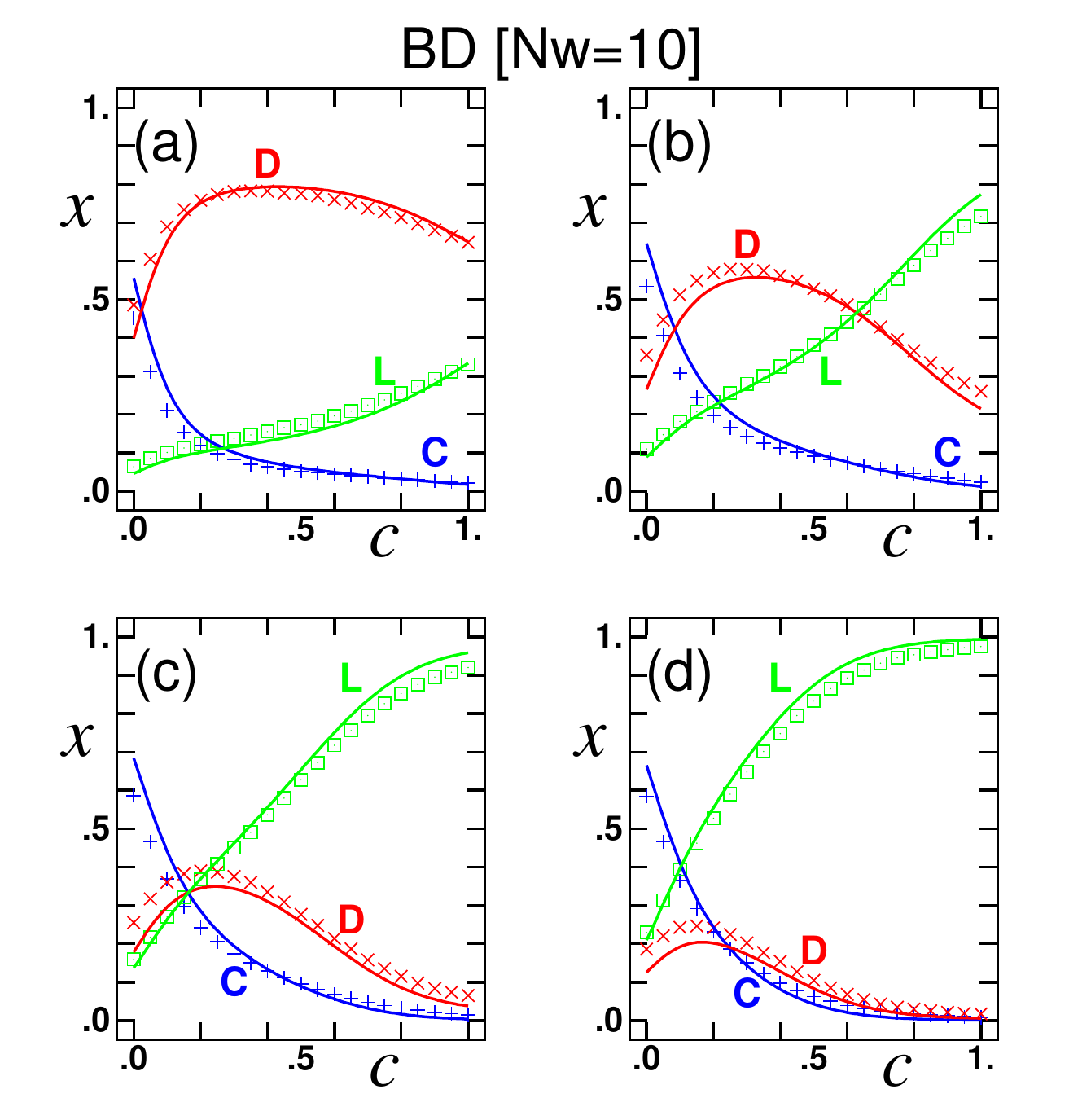}
\caption{
Abundance $\xc$, $\xd$, and $\xl$ vs. $c$ for BD with $N=50$
and $w=0.2$ for four different values of $g$, (a) 0, (b) 0.2, (c)
0.4, and (d) 0.6. Blue plus, red cross, and green square symbols 
represent the $\xc$, $\xd$, and $\xl$ respectively. Blue, red, and
green solid lines are abundance of Eq.~(\ref{e.eigen}).
Mutation rate is $u=0.0002$.
} 
\label{f.xcBD2}
\end{figure}
\begin{figure}[!h]
\includegraphics[width=.45\textwidth,keepaspectratio=true]{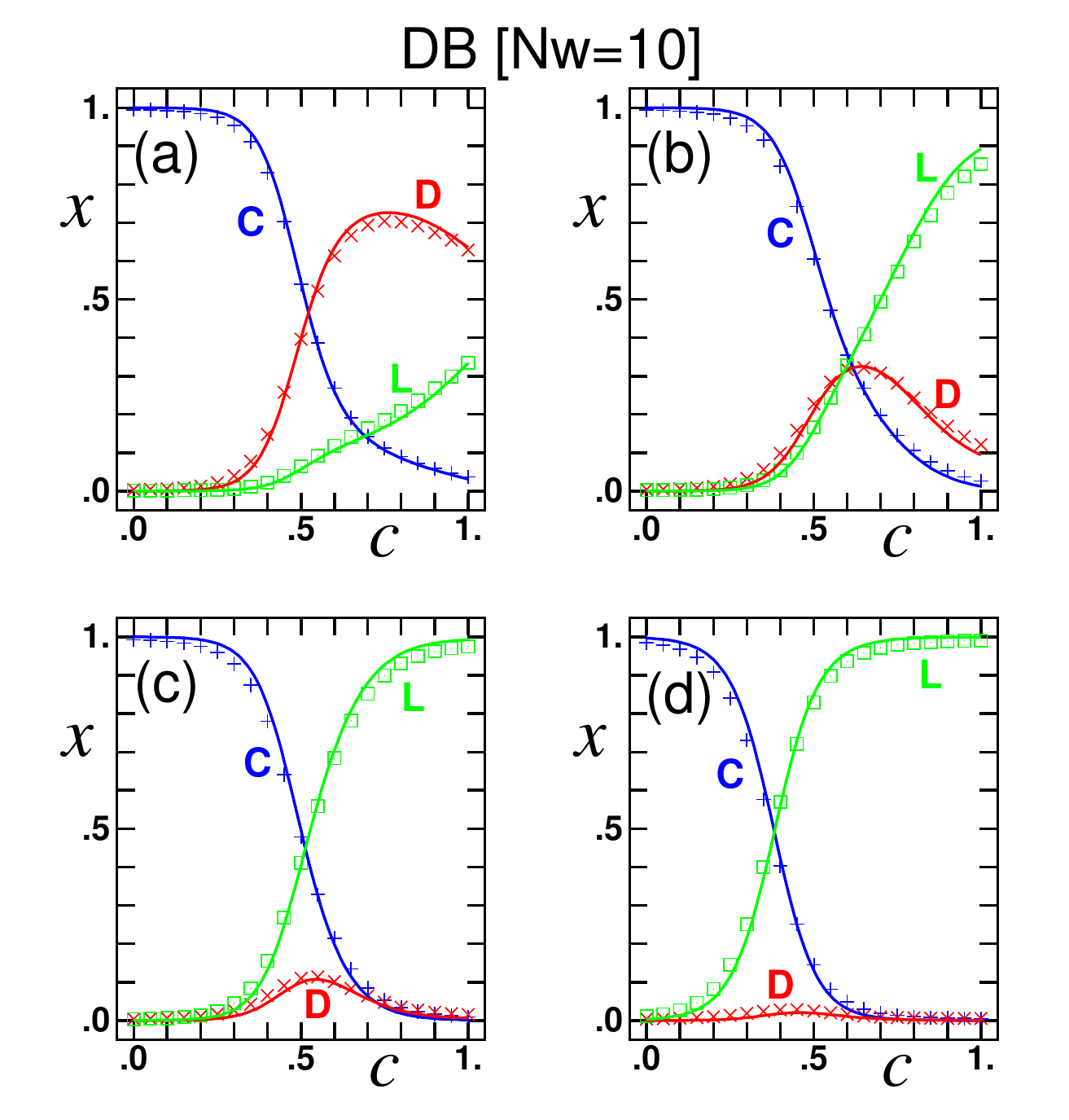}
\caption{
Abundance $\xc$, $\xd$, and $\xl$ vs. $c$ for DB with $N=50$
and $w=0.2$ for four different values of $g$, (a) 0, (b) 0.2, (c)
0.4, and (d) 0.6. Blue plus, red cross, and green square symbols 
represent the $\xc$, $\xd$, and $\xl$ respectively. Blue, red, and
green solid lines are abundance of Eq.~(\ref{e.eigen}).
Mutation rate is $u=0.0002$.
} 
\label{f.xcDB2}
\end{figure}

To get statistical properties, we perform $M=6\times10^4$ 
independent simulations and calculate the average
frequencies of strategies. 
We monitor the time evolution of the average frequencies
and see if the population evolves to a steady state
in which average frequency remains constant.
In the ensemble of steady states, 
we believe that the probability distribution of frequencies are stationary. 
For a single simulation, frequencies in the population
may oscillate through mutation-fixation 
cycles for small mutation rates.
However, the ensemble average of $M$ independent simulations effectively
provides mean frequencies equivalent to time average over many
fixations. We call this mean frequency as abundance. 

\begin{figure}[!h]
\includegraphics[width=.45\textwidth,keepaspectratio=true]{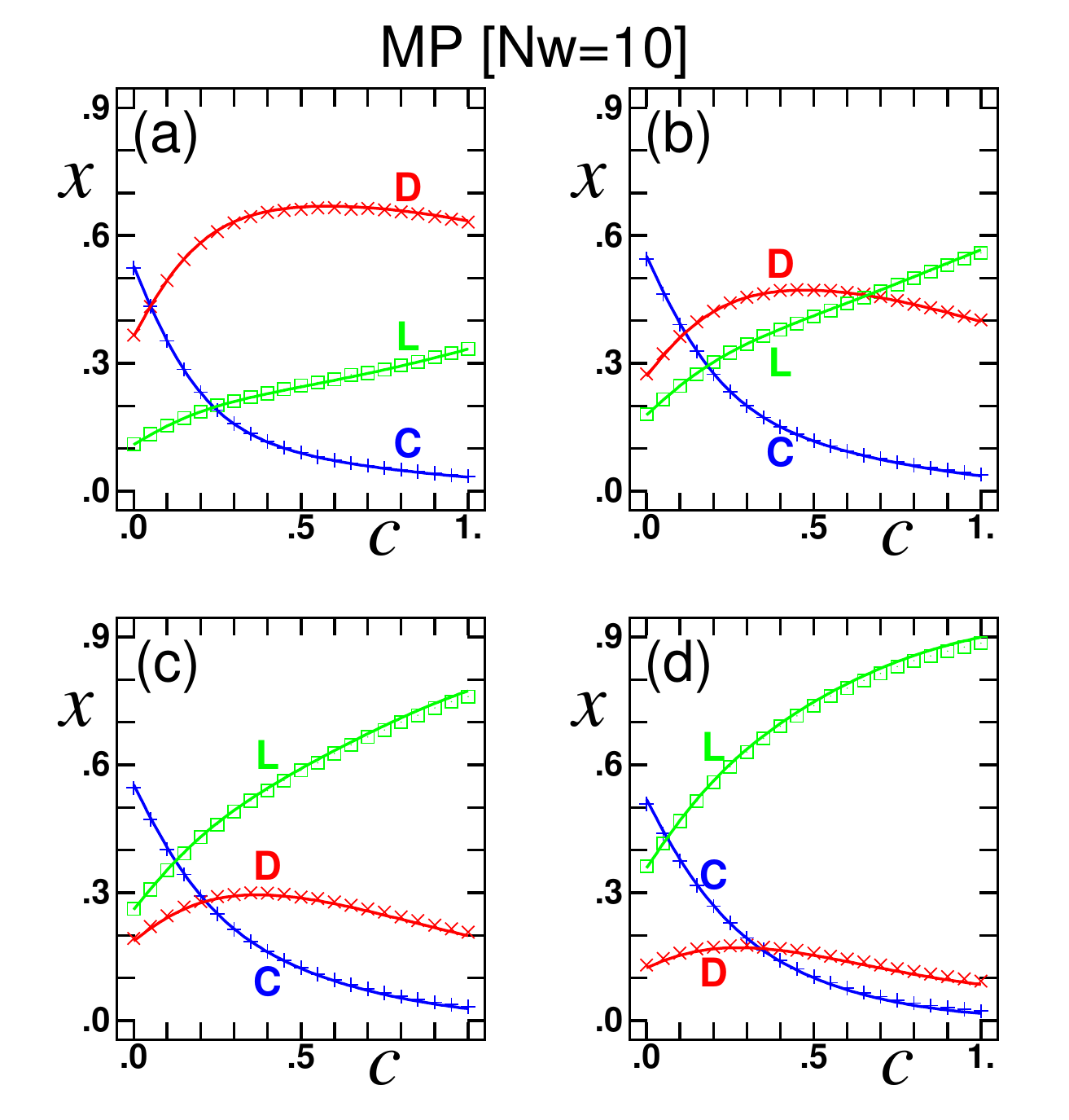}
\caption{
Abundance $\xc$, $\xd$, and $\xl$ vs. $c$ for MP with $N=50$
and $w=0.2$ for four different values of $g$, (a) 0, (b) 0.2, (c)
0.4, and (d) 0.6. Blue plus, red cross, and green square symbols 
represent the $\xc$, $\xd$, and $\xl$ respectively. Blue, red, and
green solid lines are abundance of Eq.~(\ref{e.eigen}).
Mutation rate is $u=0.0002$.
} 
\label{f.xcMP2}
\end{figure}

Time to reach a steady state from the random initial configuration 
increases rapidly with population size~$N$. Hence, we simulate 
relatively small population of $N=50$. We use mutation rate $u=0.0002$ 
such that $Nu=0.01$ in all simulations. 
 
We first measure abundance of cooperators, $\xc$, defectors, $\xd$,
and loners, $\xl$ in the small $Nw$ regime with $Nw=0.1$.  
Abundance versus cost, $x$-$c$ plots are shown 
in Fig.~\ref{f.xcBD1},~\ref{f.xcDB1}, and~\ref{f.xcMP1} for BD, DB, and
MP respectively. For each updating process, 
we simulate population dynamics with 
21 different values of~$c$, $c=0.$,\ 0.05, $\ldots$\,,\, 1,
for each of four different values of $g$, (a) 0, (b) 0.2, (c) 0.4, and (d) 0.6. 
Blue plus, red cross, and green square symbols 
represent the $\xc$, $\xd$, and $\xl$ respectively. 
They are compared with abundance of Eq.~(\ref{e.eigen}), 
calculated using fixation probabilities, which are represented by 
blue, red, and green solid lines.
We first note that the abundance of all strategies
are around 1/3 as expected in the $wN$ limit.
Measured data from simulations are consistent with
abundance of Eq.~(\ref{e.eigen}) except a
tiny but systematic deviation. When abundance is 
larger than 1/3, measured data tend to stay below
the lines while they seem to stay above the lines when 
it is smaller than 1/3. These deviations seem to come from 
the fact that we use finite mutation rate ($u=0.0002$)
instead of infinitesimal rate. Random mutations
make abundance move to the average value (1/3) regardless 
of its strategy. Except this small discrepancy, simulation data seem
to follow all features of calculated abundance of Eq.~(\ref{e.eigen}).
For example, $\xd$ and $\xl$ increase linearly   
and $\xc$ decreases linearly with increasing~$c$. 
Especially, we note that 
crossing points of $\xc$ and $\xd$ are 
independent of $g$ as predicted. 
$\xc$ and $\xd$ meet near $c=1/5$ for BD and MP,
and near $c=7/11\simeq 0.64$ for DB.

\begin{figure}[!h]
\includegraphics[width=.3\textwidth,keepaspectratio=true]{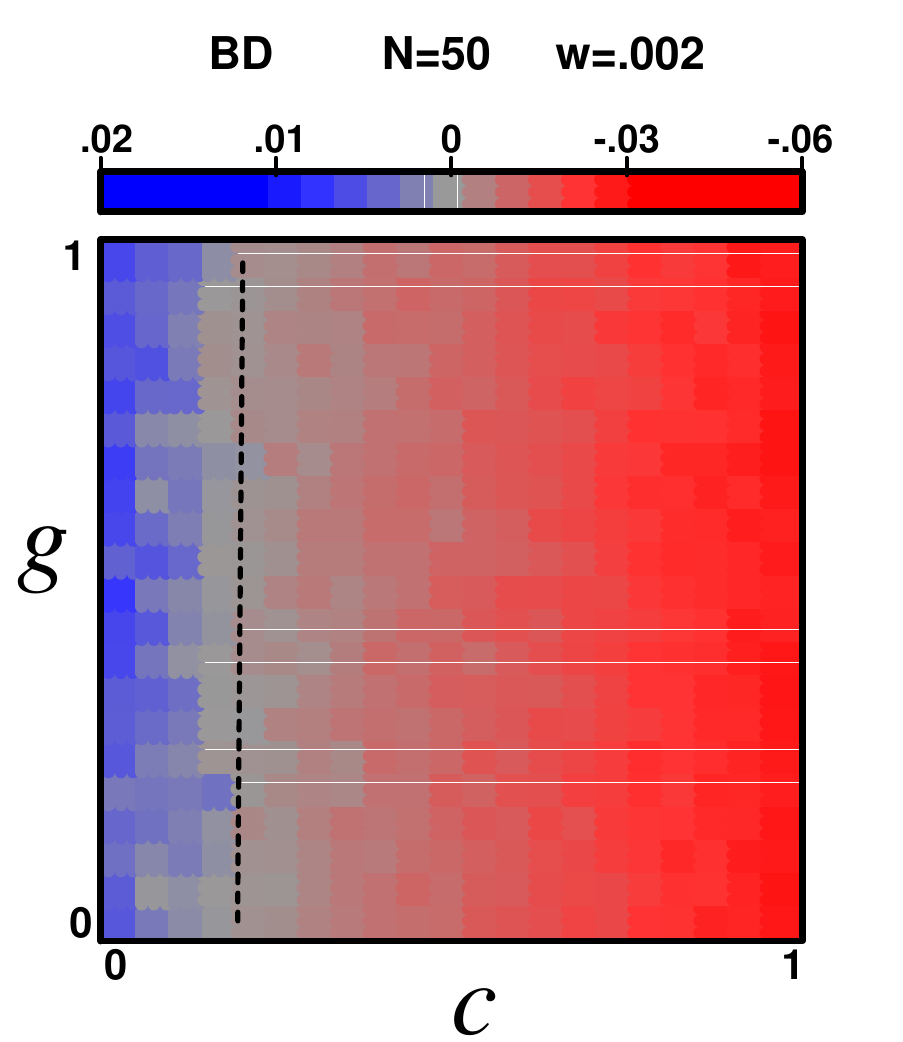}
\includegraphics[width=.3\textwidth,keepaspectratio=true]{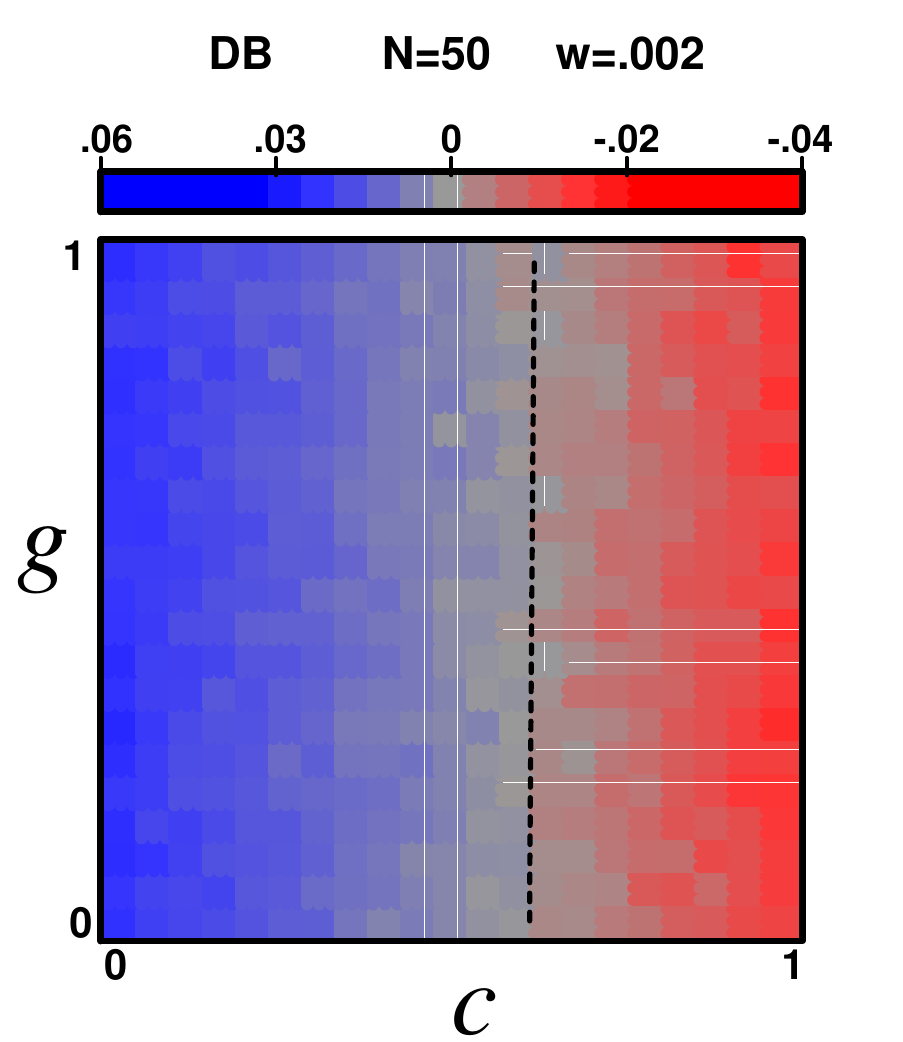}
\includegraphics[width=.3\textwidth,keepaspectratio=true]{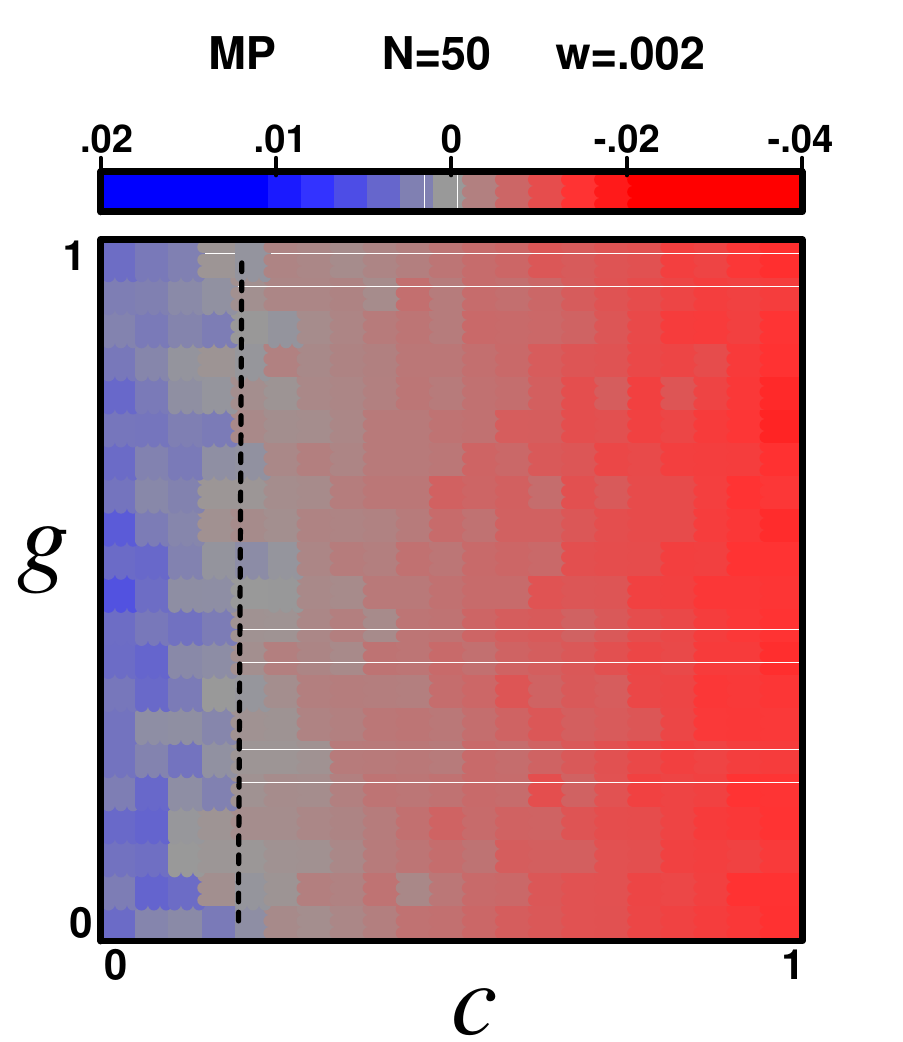}
\caption{
Normalized abundance difference between cooperators and defectors,
$r=(\xc-\xd)/(\xc+\xd)$ in the small $Nw$ regime with $N=50$ and 
$w=0.002$ ($Nw=0.1$), for (a) BD, (b) DB, and (c) MP. 
Mutation rate is $u=0.0002$. The vertical blue and the
horizontal red paintings represent C-rich and D-rich regions
respectively. The dashed line is the boundary for $\xc=\xd$ 
in the low mutation limit of $u\rightarrow 0$.
} 
\label{f.mu.ph1}
\end{figure}

Simulation data for the large $Nw$ also follow the predicted abundance of 
Eq.~(\ref{e.eigen}) quite well. 
Figures~\ref{f.xcBD2}, \ref{f.xcDB2} and and~\ref{f.xcMP2} show  
$x$-$c$ plots for BD, DB, and MP respectively for $w=0.2$ ($Nw=10$). 
As before,  $\xc$, $\xd$, and $\xl$ versus $c$ graphs 
are represented by blue plus, red cross, and green square symbols 
respectively for four different values of $g$, (a) 0, (b) 0.2,
(c) 0.4, and (d) 0.6. They are compared with calculated 
abundance of Eq.~(\ref{e.eigen}), shown by blue, red, and green solid lines. 
As before, we observe small but
systematic discrepancies between simulation data and 
predicted abundance of Eq.~(\ref{e.eigen}).
Measure abundance deference between (different) strategies 
are smaller than the predictions. 
This can be understood from the fact that mutations reduce the abundance difference
between strategies. Aside from this systematic deviation,
simulation data follow the features of predicted abundance very well. 

\begin{figure}[!h]
\includegraphics[width=.3\textwidth,keepaspectratio=true]{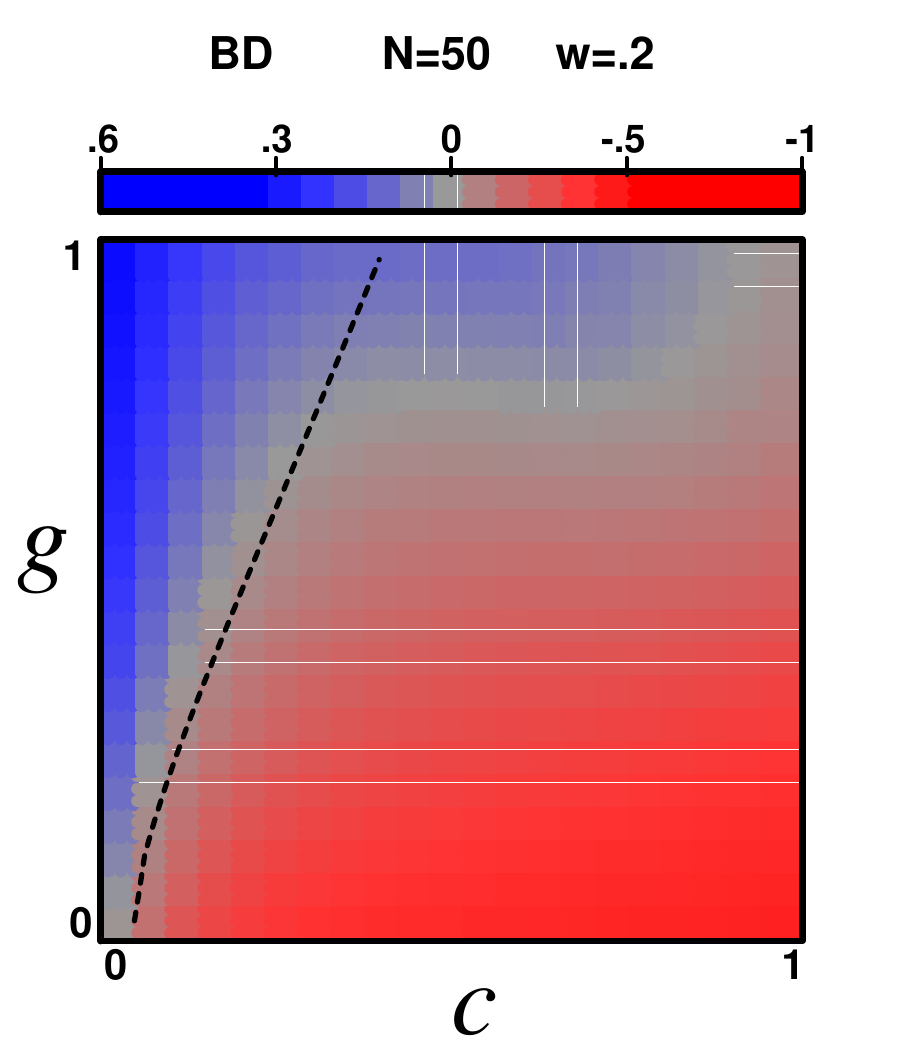}
\includegraphics[width=.3\textwidth,keepaspectratio=true]{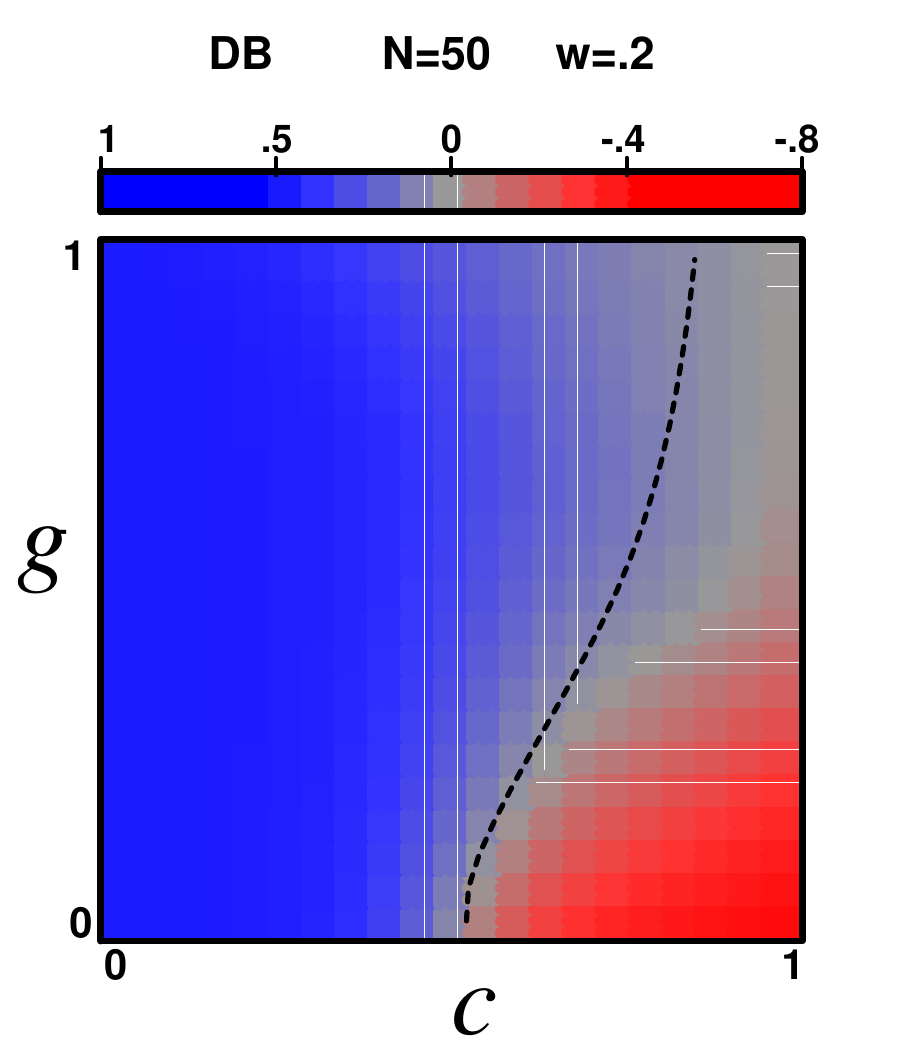}
\includegraphics[width=.3\textwidth,keepaspectratio=true]{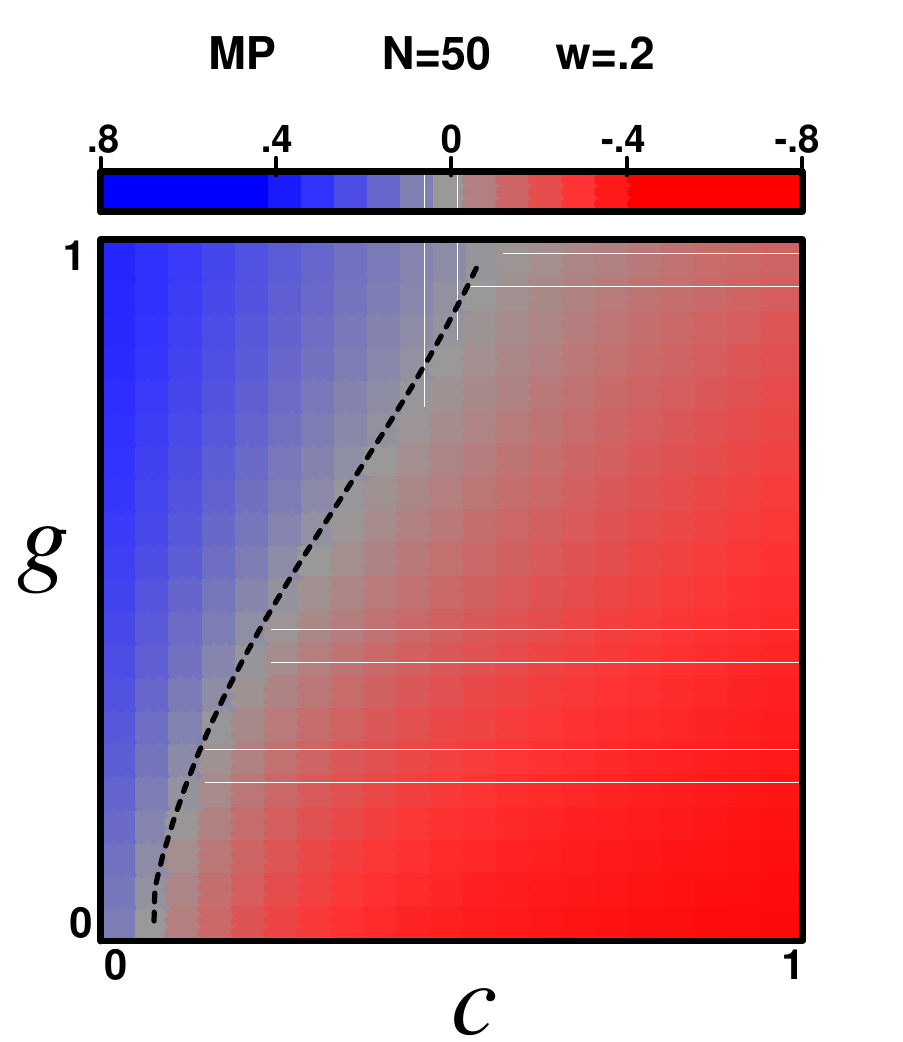}
\caption{
Normalized abundance difference between cooperators and defectors,
$r=(\xc-\xd)/(\xc+\xd)$ in the large $Nw$ regime with 
$N=50$ and $w=0.2$ ($Nw=10$), for (a) BD, (b) DB, and (c) MP. 
Mutation rate is $u=0.0002$. The vertical blue and the
horizontal red paintings represent C-rich and D-rich regions
respectively. The dashed line is the boundary for $\xc=\xd$ 
in the low mutation limit. 
} 
\label{f.mu.ph2}
\end{figure}

We now investigate C-rich and D-rich regions in the parameter space of
$c$ and~$g$ and compare them with those in the low mutation limit.
We first measure $\xc$ and $\xd$ for $21\times21$ different $c$-$g$
pairs in $r\in[0\ 1]$ and $g\in[0\ 1]$ with intervals of 0.05.
Then, we plot a normalized abundance difference between cooperators
and defectors, $r=(\xc-\xd)/(\xc+\xd)$ in color in $21\times 21$ mesh
in the  $c$-$g$ parameter space (Figs.~\ref{f.mu.ph1} and~\ref{f.mu.ph2})
to illustrate C-rich and D-rich regions. As before, we use population
of $N=50$ with mutation rate  $u=0.0002$. 
The blue-vertical and the red-horizontal paintings represent C-rich
and D-rich regions respectively.

Figure~\ref{f.mu.ph1} shows the normalized abundance difference, 
$r$ in the small $Nw$ regime for the three processes
with $w=0.002$ ($Nw=0.1$). 
As predicted by the panels~(a) in Fig.~1-3, 
blue-rich region changes to red-rich region as $c$
increases, more or less, uniformly regardless of $g$ values. 
The phase boundaries calculated in the low mutation limit 
are shown in black-dashed lines. 
Those lines locate near $c=1/5$ for BD and MP and 
near $c=7/11$ for DB updating and they are consistent 
to the boundaries between two colors. 

Boundaries (of C-rich and D-rich regions) obtained from the 
simulations for the large $Nw$ regime are also consistent 
with those calculated in the low mutation limit.
Figure~\ref{f.mu.ph2} shows the normalized abundance difference, 
$r$ in color for the three processes with $w=0.2$ ($Nw=10$)
in the $c$-$g$ parameter space. As in Fig.~\ref{f.mu.ph1},
the blue-vertical and the red-horizontal paintings represent C-rich
and D-rich regions respectively.  
The phase boundaries calculated in the low mutation limit are shown 
in black-dashed lines. They are consistent with color boundaries quite
well expect for large~$g$ for BD updating. 
We observe that cooperators favored over defectors for wider range
of~$c$ for large~$g$ for BD updating. However, the absolute abundance
of cooperators is small (although it is still larger than $\xd$) when $g$
is large, since loners prevail the population.   

\section{Conclusion}
We have analyzed strategy selection in optional games on cycles and on 
complete graphs and found a non-trivial interaction between
volunteering and spatial selection.   

For $2\times 2$ games on cycles using exponential fitness, we have
presented a closed form expression for the fixation probability for
any intensity of selection and any population size. Using this
fixation probability, we have found the conditions for strategy
selection analytically in the limits of weak intensity of selection
and large population size. We have presented results for two orders of
limits: (i) $w\to 0$ followed by $N\to \infty$ (which we call the
$wN$-limit) and (ii)  $N\to \infty$ followed by $w\to 0$ (which we
call the $Nw$-limit). In the first case we have $wN \ll 1$; in the second
we have $Nw \gg 1$. We have also obtained numerical results for
finite $w$ in the low mutation limit.  

According to our observations, increasing the number of loner
strategies relaxes the social dilemma and promotes evolution of
cooperation. Increasing the number of loner strategies is 
equivalent to increasing mutational bias toward loner strategies. 
More loner strategies (or equivalently, more bias in mutation toward
loners) favors cooperation by enabling loners to invade defector clusters and facilitate the return of cooperators.
In the limit of an infinite number of loner strategies the social
dilemma is completely resolved for any selection intensity.   
For high intensity of selection ($w\gg 1$), the social dilemma can be
fully resolved if there is mutational bias toward loner strategies
(or there are more than one loner strategies). 

While optionality of the game and spatial population structure both
support cooperation, we have not found evidence of synergy between
these mechanisms.  This lack of synergy appears due to the fact that
these mechanisms act in  different ways.  Spatial structure supports
cooperation by allowing cooperators to isolate themselves, while
optionality supports cooperation by allowing loners to infiltrate
defectors.  Neither mechanism appears to improve the efficacy of the
other.  In fact, for strong selection (the $w \to \infty$ limit) these
mechanisms appear to counteract one another, in that there are
parameter combinations for which coopeation is favored in the
well-mixed population but disfavored for DB updating on the cycle. 

We speculate that the role of loner strategies in relaxing social
dilemmas, which we observe in our study, is qualitatively valid for
games on general graphs. Since the population structures in our study,
cycles and complete graphs, are at the two extreme ends of the
spectrum of spatial structures, we expect loner strategies in optional
games on other graphs also to relax social dilemma. 
The relaxation effect of volunteering increases as more loner strategies are
available.  

\section{Acknowledgments}
Support from the program for Foundational Questions in Evolutionary
Biology (FQEB), the National Philanthropic Trust, the John Templeton 
Foundation and the National Research Foundation of Korea grant
(NRF-2010-0022474) is gratefully acknowledged.


\newpage


\begin{thebibliography}{}

\bibitem[Allen {\em et~al.}, 2013]{Allen:2013vu}
Allen, B., Gore, J., Nowak, M.~A.  \& Bergstrom, C.~T. 2013{\em{}}.
\newblock {Spatial dilemmas of diffusible public goods}.
\newblock eLife,  2, e01169.

\bibitem[Allen \& Nowak, 2014]{Allen:2014bu}
Allen, B. \& Nowak, M. 2014{\em{}}.
\newblock {Games on graphs}.
\newblock EMS Surv. Math. Sci. 1 (1), 113--151.

\bibitem[Batali \& Kitcher, 1995]{Batali:1995uz}
Batali, J. \& Kitcher, P. 1995{\em{}}.
\newblock {Evolution of altriusm in optional and compulsory games}.
\newblock J Theor Biol,  175, 161--171.

\bibitem[Boyd \& Richerson, 1992]{Boyd:1992dt}
Boyd, R. \& Richerson, P.~J. 1992{\em{}}.
\newblock {Punishment allows the evolution of cooperation (or anything else) in
  sizable groups}.
\newblock Ethology and Sociobiology,  13 (3), 171--195.

\bibitem[Challet \& Zhang, 1997]{Challet:1997wz}
Challet, D. \& Zhang, Y.-C. 1997{\em{}}.
\newblock {Emergence of Cooperation and Organization in an Evolutionary Game}.
\newblock Physica A-Statistical Mechanics and Its Applications,  246, 407--418.

\bibitem[Cressman, 2003]{Cressman:2003ux}
Cressman, R. 2003{\em{}}.
\newblock {\em {Evolutionary Dynamics and Extensive Form Games}}.
\newblock MIT Press, Cambridge.

\bibitem[De~Silva {\em et~al.}, 2009]{DeSilva:2010uz}
De~Silva, H., Hauert, C., Traulsen, A.  \& Sigmund, K. 2009{\em{}}.
\newblock {Freedom, enforcement, and the social dilemma of strong altruism}.
\newblock J Evol Econ,  20 (2), 203--217.

\bibitem[Foster \& Young, 1990]{Foster:1990io}
Foster, D. \& Young, P. 1990{\em{}}.
\newblock {Stochastic evolutionary game dynamics}.
\newblock Theor Popul Biol,  38 (2), 219--232.

\bibitem[Friedman, 1998]{Friedman:1998uy}
Friedman, D. 1998{\em{}}.
\newblock {On economic applications of evolutionary game theory}.
\newblock J Evol Econ,  8 (1), 15--43.

\bibitem[Fu {\em et~al.}, 2007{\em{a}}]{Fu:2007vo}
Fu, F., Chen, X., Liu, L.  \& Wang, L. 2007{\em{a}}.
\newblock {Social dilemmas in an online social network: The structure and
  evolution of cooperation}.
\newblock Phys Lett A,  371 (1-2), 58--64.

\bibitem[Fu {\em et~al.}, 2007{\em{b}}]{Fu:2007we}
Fu, F., Chen, X., Liu, L.  \& Wang, L. 2007{\em{b}}.
\newblock {Promotion of cooperation induced by the interplay between structure
  and game dynamics}.
\newblock Physica A-Statistical Mechanics and Its Applications,  383 (2),
  651--659.

\bibitem[Garcia \& Traulsen, 2012]{Garcia:2012cg}
Garcia, J. \& Traulsen, A. 2012{\em{}}.
\newblock {The structure of mutations and the evolution of cooperation}.
\newblock Plos One,  7 (4), e35287.

\bibitem[Gokhale \& Traulsen, 2011]{Gokhale:2011hk}
Gokhale, C.~S. \& Traulsen, A. 2011{\em{}}.
\newblock {Strategy abundance in evolutionary many-player games with multiple
  strategies}.
\newblock J Theor Biol,  283 (1), 180--191.

\bibitem[Hauert, 2002]{Hauert:2002ua}
Hauert, C. 2002{\em{}}.
\newblock {Volunteering as Red Queen Mechanism for Cooperation in Public Goods
  Games}.
\newblock Science,  296 (5570), 1129--1132.

\bibitem[Hauert {\em et~al.}, 2002]{Hauert:2002hd}
Hauert, C., De~Monte, S., Hofbauer, J.  \& Sigmund, K. 2002{\em{}}.
\newblock {Replicator dynamics for optional public good games}.
\newblock J Theor Biol,  218 (2), 187--194.

\bibitem[Hauert {\em et~al.}, 2014]{Debarre:2014tq}
Hauert, C., Doebeli, M.  \& barre, F. D.~e. 2014{\em{}}.
\newblock {Social evolution in structured populations}.
\newblock Nat Commun,  5, 3409.

\bibitem[Hauert {\em et~al.}, 2006]{Hauert:2006kz}
Hauert, C., Michor, F., Nowak, M.~A.  \& Doebeli, M. 2006{\em{}}.
\newblock {Synergy and discounting of cooperation in social dilemmas}.
\newblock J Theor Biol,  239 (2), 195--202.

\bibitem[Hauert {\em et~al.}, 2007]{Hauert:2007js}
Hauert, C., Traulsen, A., Brandt, H., Nowak, M.~A.  \& Sigmund, K. 2007{\em{}}.
\newblock {Via Freedom to Coercion: The Emergence of Costly Punishment}.
\newblock Science,  316 (5833), 1905--1907.

\bibitem[Hilbe \& Sigmund, 2010]{Hilbe:2010vs}
Hilbe, C. \& Sigmund, K. 2010{\em{}}.
\newblock {Incentives and opportunism: from the carrot to the stick}.
\newblock P R Soc B,  277 (1693), 2427--2433.

\bibitem[Hofbauer \& Sigmund, 1998]{Hofbauer:1998wn}
Hofbauer, J. \& Sigmund, K. 1998{\em{}}.
\newblock {Evolutionary Games and Population Dynamics}.
\newblock Cambridge University Press.

\bibitem[Hofbauer \& Sigmund, 1988]{Hofbauer:1988ve}
Hofbauer, J. \& Sigmund, K.~S. 1988{\em{}}.
\newblock {\em {The theory of evolution and dynamical systems}}.
\newblock Cambridge University Press.

\bibitem[Imhof \& Nowak, 2006]{Imhof:2006ju}
Imhof, L.~A. \& Nowak, M.~A. 2006{\em{}}.
\newblock {Evolutionary game dynamics in a Wright-Fisher process}.
\newblock J. Math. Biol. 52 (5), 667--681.

\bibitem[Kitcher, 1993]{Kitcher:1993tc}
Kitcher, P. 1993{\em{}}.
\newblock {The evolution of human altruism}.
\newblock The Journal of Philosophy,  90 (10), 497--516.

\bibitem[Lieberman {\em et~al.}, 2005]{Lieberman:2005tn}
Lieberman, E., Hauert, C.  \& Nowak, M.~A. 2005{\em{}}.
\newblock {Evolutionary dynamics on graphs}.
\newblock Nature,  433 (7023), 312--316.

\bibitem[Maciejewski, 2014]{Maciejewski:2014ko}
Maciejewski, W. 2014{\em{}}.
\newblock {Reproductive value in graph-structured populations}.
\newblock J Theor Biol,  340, 285--293.

\bibitem[Michor \& Nowak, 2002]{Michor:2002wu}
Michor, F. \& Nowak, M.~A. 2002{\em{}}.
\newblock {Evolution: The good, the bad and the lonely}.
\newblock Nature,  419 (6908), 677--679.

\bibitem[Nakamaru \& Iwasa, 2005]{Nakamaru:2005ug}
Nakamaru, M. \& Iwasa, Y. 2005{\em{}}.
\newblock {The evolution of altruism by costly punishment in lattice-structured
  populations: score-dependent viability versus score-dependent fertility}.
\newblock Evolutionary ecology research,  7, 853--870.

\bibitem[Nakamaru {\em et~al.}, 1997]{Nakamaru:1997uq}
Nakamaru, M., Matsuda, H.  \& Iwasa, Y. 1997{\em{}}.
\newblock {The Evolution of Cooperation in a Lattice-Structured Population}.
\newblock J Theor Biol,  184 (1), 65--81.

\bibitem[Nowak, 2004]{Nowak:2004vt}
Nowak, M.~A. 2004{\em{}}.
\newblock {Evolutionary Dynamics of Biological Games}.
\newblock Science,  303 (5659), 793--799.

\bibitem[Nowak, 2006{\em{a}}]{Nowak:2006bt}
Nowak, M.~A. 2006{\em{a}}.
\newblock {Five Rules for the Evolution of Cooperation}.
\newblock Science,  314 (5805), 1560--1563.

\bibitem[Nowak, 2006{\em{b}}]{Nowak:2006vy}
Nowak, M.~A. 2006{\em{b}}.
\newblock {\em {Evolutionary Dynamics: Exploring the Equations of Life}}.
\newblock Harvard University Press, Cambridge.

\bibitem[Nowak, 2012]{Nowak:2012cw}
Nowak, M.~A. 2012{\em{}}.
\newblock {Evolving cooperation}.
\newblock J Theor Biol,  299, 1--8.

\bibitem[Nowak {\em et~al.}, 1994]{Nowak:1994vj}
Nowak, M.~A., Bonhoeffer, S.  \& May, R.~M. 1994{\em{}}.
\newblock {Spatial games and the maintenance of cooperation}.
\newblock Proceedings of the National Academy of Sciences,  91, 4877--4811.

\bibitem[Nowak \& May, 1992]{Nowak:1992vx}
Nowak, M.~A. \& May, R.~M. 1992{\em{}}.
\newblock {Evolutionary games and spatial chaos}.
\newblock Nature,  359, 826--829.

\bibitem[Nowak {\em et~al.}, 2004]{Nowak:2004td}
Nowak, M.~A., Sasaki, A., Taylor, C.  \& Fudenberg, D. 2004{\em{}}.
\newblock {Emergence of cooperation and evolutionary stability in finite
  populations}.
\newblock Nature,  428 (6983), 646--650.

\bibitem[Nowak {\em et~al.}, 2010]{Nowak:2010ik}
Nowak, M.~A., Tarnita, C.~E.  \& Antal, T. 2010{\em{}}.
\newblock {Evolutionary dynamics in structured populations}.
\newblock Philosophical Transactions of the Royal Society B: Biological
  Sciences,  365 (1537), 19--30.

\bibitem[Ohtsuki {\em et~al.}, 2006]{Ohtsuki:2006ug}
Ohtsuki, H., Hauert, C., Lieberman, E.  \& Nowak, M.~A. 2006{\em{}}.
\newblock {A simple rule for the evolution of cooperation on graphs and social
  networks}.
\newblock Nature,  441 (25), 502--505.

\bibitem[Ohtsuki \& Nowak, 2006]{Ohtsuki:2006gp}
Ohtsuki, H. \& Nowak, M.~A. 2006{\em{}}.
\newblock {Evolutionary games on cycles}.
\newblock P R Soc B,  273 (1598), 2249--2256.

\bibitem[Perc, 2011]{Perc:2011th}
Perc, M. 2011{\em{}}.
\newblock {Does strong heterogeneity promote cooperation by group
  interactions?}
\newblock New J. Phys. 13 (12), 123027.

\bibitem[Perc \& Szolnoki, 2010]{Perc:2010tr}
Perc, M. \& Szolnoki, A. 2010{\em{}}.
\newblock {Coevolutionary games---A mini review}.
\newblock Biosystems,  99 (2), 109--125.

\bibitem[Rand \& Nowak, 2011]{Rand:2011gy}
Rand, D.~G. \& Nowak, M.~A. 2011{\em{}}.
\newblock {The evolution of antisocial punishment in optional public goods
  games}.
\newblock Nat Commun,  2, 434.

\bibitem[Rand \& Nowak, 2013]{Rand:2013vx}
Rand, D.~G. \& Nowak, M.~A. 2013{\em{}}.
\newblock {Human cooperation}.
\newblock Trends in cognitive sciences,  17 (8), 413--425.

\bibitem[Santos \& Pacheco, 2005]{Santos:2005im}
Santos, F. \& Pacheco, J. 2005{\em{}}.
\newblock {Scale-Free Networks Provide a Unifying Framework for the Emergence
  of Cooperation}.
\newblock Phys. Rev. Lett. 95 (9), 098104.

\bibitem[Santos {\em et~al.}, 2008]{Santos:2008tg}
Santos, F.~C., Santos, M.~D.  \& Pacheco, J.~M. 2008{\em{}}.
\newblock {Social diversity promotes the emergence of cooperation in public
  goods games}.
\newblock Nature,  454 (7201), 213--216.

\bibitem[Sigmund, 2007]{Sigmund:2007vz}
Sigmund, K. 2007{\em{}}.
\newblock {Punish or perish? Retaliation and collaboration among humans}.
\newblock Trends in Ecology {\&} Evolution,  22 (11), 593--600.

\bibitem[Szab{\'o} \& F{\'a}th, 2007]{Szabo:2007uy}
Szab{\'o}, G. \& F{\'a}th, G. 2007{\em{}}.
\newblock {Evolutionary games on graphs}.
\newblock Physics Reports,  446 (4-6), 97--216.

\bibitem[Szab{\'o} \& Hauert, 2002{\em{a}}]{Szabo:2002cv}
Szab{\'o}, G. \& Hauert, C. 2002{\em{a}}.
\newblock {Evolutionary prisoner's dilemma games with voluntary participation}.
\newblock Phys. Rev. E,  66 (6), 062903.

\bibitem[Szab{\'o} \& Hauert, 2002{\em{b}}]{Szabo:2002gd}
Szab{\'o}, G. \& Hauert, C. 2002{\em{b}}.
\newblock {Phase transitions and volunteering in spatial public goods games}.
\newblock Phys. Rev. Lett. 89 (11), 118101.

\bibitem[Tarnita {\em et~al.}, 2009{\em{a}}]{Tarnita:2009vna}
Tarnita, C.~E., Antal, T., Ohtsuki, H.  \& Nowak, M.~A. 2009{\em{a}}.
\newblock {Evolutionary dynamics in set structured populations}.
\newblock Proceedings of the National Academy of Sciences,  106 (21),
  8601--8604.

\bibitem[Tarnita {\em et~al.}, 2009{\em{b}}]{Tarnita:2009vn}
Tarnita, C.~E., Ohtsuki, H., Antal, T., Fu, F.  \& Nowak, M.~A. 2009{\em{b}}.
\newblock {Strategy selection in structured populations}.
\newblock J Theor Biol,  259 (3), 570--581.

\bibitem[Tarnita {\em et~al.}, 2011]{Tarnita:2011tf}
Tarnita, C.~E., Wage, N.  \& Nowak, M.~A. 2011{\em{}}.
\newblock {Multiple strategies in structured populations}.
\newblock Proceedings of the National Academy of Sciences,  108 (6),
  2334--2337.

\bibitem[Taylor {\em et~al.}, 2004]{Fudenberg:2004ub}
Taylor, C., Fudenberg, D., Sasaki, A.  \& Nowak, M.~A. 2004{\em{}}.
\newblock {Evolutionary game dynamics in finite populations}.
\newblock Bull. Math. Biol. 66 (6), 1621--1644.

\bibitem[Traulsen {\em et~al.}, 2009]{Traulsen:2009cg}
Traulsen, A., Hauert, C., De~Silva, H., Nowak, M.~A.  \& Sigmund, K.
  2009{\em{}}.
\newblock {Exploration dynamics in evolutionary games}.
\newblock Proceedings of the National Academy of Sciences,  106 (3), 709--712.

\bibitem[Traulsen {\em et~al.}, 2006]{Traulsen:2006uf}
Traulsen, A., Pacheco, J.~M.  \& Imhof, L.~A. 2006{\em{}}.
\newblock {Stochasticity and evolutionary stability}.
\newblock Phys. Rev. E,  74, 021905.

\bibitem[Traulsen {\em et~al.}, 2008]{Traulsen:2008ee}
Traulsen, A., Shoresh, N.  \& Nowak, M.~A. 2008{\em{}}.
\newblock {Analytical Results for Individual and Group Selection of Any
  Intensity}.
\newblock Bull. Math. Biol. 70 (5), 1410--1424.

\bibitem[van Veelen \& Nowak, 2012]{vanVeelen:2012wl}
van Veelen, M. \& Nowak, M.~A. 2012{\em{}}.
\newblock {Multi-player games on the cycle}.
\newblock J Theor Biol,  292, 116--128.

\bibitem[Vincent \& Brown, 2005]{Vincent:2005wo}
Vincent, T.~L. \& Brown, J.~S. 2005{\em{}}.
\newblock {\em {Evolutionary Game Theory, Natural Selection, and Darwinian
  Dynamics}}.
\newblock Cambridge University Press.

\bibitem[Weibull, 1997]{Weibull:1997vv}
Weibull, J.~W. 1997{\em{}}.
\newblock {\em {Evolutionary Game Theory}}.
\newblock MIT Press.

\bibitem[Wu {\em et~al.}, 2013]{Wu:2013uh}
Wu, B., Garcia, J., Hauert, C.  \& Traulsen, A. 2013{\em{}}.
\newblock {Extrapolating weak selection in evolutionary games}.
\newblock PLoS Comp Biol,  9 (12), e1003381.

\end{thebibliography}
\end{document}